\documentclass[lettersize,journal]{IEEEtran}
\usepackage{amsmath,amsfonts}
\usepackage{algorithmic}
\usepackage{algorithm}
\usepackage{array}
\usepackage[caption=false,font=normalsize,labelfont=sf,textfont=sf]{subfig}
\usepackage{textcomp}
\usepackage{stfloats}
\usepackage{url}
\usepackage{verbatim}
\usepackage{graphicx}
\usepackage{cite}
\usepackage{bm}
\usepackage{makecell}
\usepackage{verbatim}
\begin{document}

\title{Spatial sparse sampling-based iterative optimization framework for GNSS Direct Position Estimation}

\author{Wei Gao, Rong Yang,~\IEEEmembership{Senior Member,~IEEE,} Jihong Huang, Xingqun Zhan,~\IEEEmembership{Senior Member,~IEEE,} and Yonggang Zhang,~\IEEEmembership{Senior Member,~IEEE,}

\thanks{This research was supported by the National Natural Science Foundation of China (U2570201). This work has been submitted to the IEEE for possible publication. {{\itshape (Corresponding author: Rong Yang.)}}}
\thanks{Manuscript received XXXXX 00, 0000; revised XXXXX 00, 0000.}}

\markboth{Journal of \LaTeX\ Class Files,~Vol.~00, No.~0, XXXXX~2025}
{Shell \MakeLowercase{\textit{et al.}}: Spatial sparse sampling-based iterative optimization framework for GNSS Direct Position Estimation: A Correlation Domain Perspective}


\maketitle

\begin{abstract}
Direct position estimation (DPE), a promising technique in Global Navigation Satellite Systems (GNSS) receivers, enables estimation of position, velocity, and time (PVT) solutions directly from correlator outputs. The conventional grid search (GS)-based DPE is computationally intensive, as it relies solely on locating the peak of the cross ambiguity function (CAF), and it does not fully leverage the PVT information present in the correlation values. This paper proposes an iterative optimization DPE framework that capitalizes on spatial coherence and spatial gradient via spatial sparse sampling (SS). In SS-DPE framework, the correlation outputs of spatial sampled PVT points all serve as measurements, where the spatial gradient and spatial coherence are derived to capture the information density and diversity of different correlation values. The analytical Cramér-Rao Bound (CRB) is derived and indicates that both spatial coherence and gradient determine the theoretical performance limit via the noise covariance and Jacobian matrices analysis. The proposed theoretical framework not only validates the feasibility of sparse sampling but also guides weight optimization to distinct correlation values, effectively integrating these insights with a general gradient-based optimization. Theoretical derivations are validated via Monte Carlo simulations. Field experiments further demonstrate the practical feasibility of the proposed SS-DPE optimization framework. Comparative analysis shows that the proposed SS-DPE achieves comparable PVT estimation accuracy to the conventional GS-DPE while consumes only sparsely sampled correlation values, improving the information utilization efficiency and reducing the computation load.
\end{abstract}


\section{Introduction}

\IEEEPARstart{T}{he} estimation of position, velocity and time (PVT) acts as the most fundamental function of Global Navigation Satellite Systems (GNSS) \cite{MLE in GNSS} - \cite{DOA estimation}. The direct position estimation (DPE) receiver is a popular technology that enables direct estimation of PVT parameters from the GNSS signal domain \cite{DPE review}. The DPE receiver offers better theoretical performance \cite{DPE Cramer–Rao Bound} and demonstrates strong robustness in urban challenge environment \cite{DPE Urban test}. The concept of DPE has also been applied into the fields such as narrowband radio frequency transmitters \cite{DPD radio narrowband} and multiple radio signals \cite{DPD Performance} -\cite{DPD NLOS doppler} .

\begin{figure}[h]
    \centering
    \includegraphics[width=0.8\linewidth]{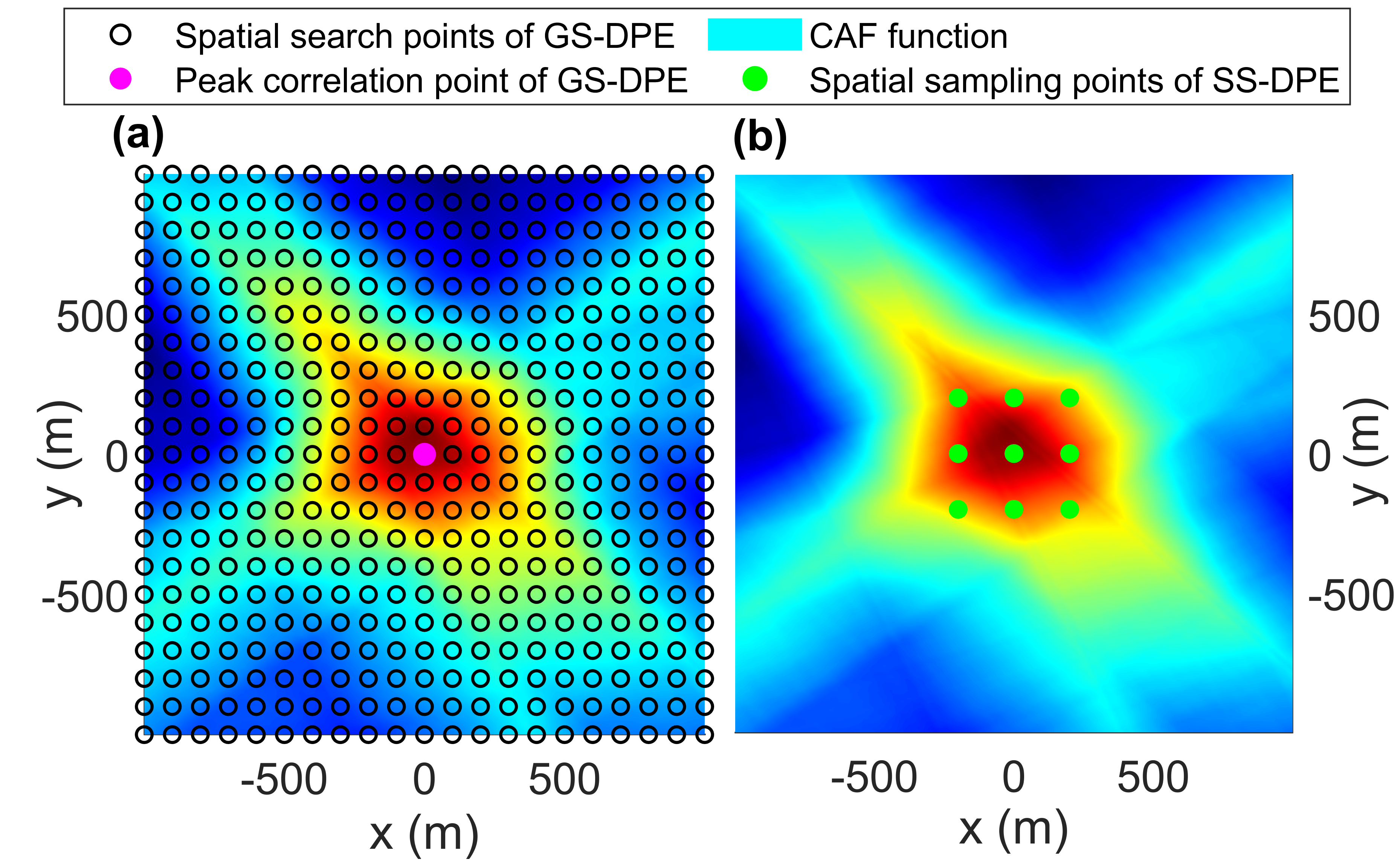}
    \caption{The principle comparison between (a) GS-DPE with large amount of spatial search points, and (b) SS-DPE with sparsely sampled spatial points.}
    \label{fig:grid_opt_demo}
\end{figure}

The current DPE receiver uses the cross ambiguity function (CAF) maximization criterion as its theoretical optimization principle  \cite{DPE definition}, which essentially maximizes the correlation values across all combined satellite channels. Under this criterion, conventional gradient-based optimization schemes are difficult to apply \cite{Iteration DPE problem}, making grid search optimization the commonly adopted solution  \cite{Parallelized_DPE}. This grid search DPE receiver (GS-DPE) exhaustively explores the possible PVT solutions, selecting PVT parameters that only maximize the CAF as the final estimation, as shown in Fig.\ref{fig:grid_opt_demo}(a).

To ensure both robustness and accuracy simultaneously, the grid search range must be large enough to encompass the CAF peak, while the search step size should also be fine enough. This, in turn, creates a fundamental dilemma: the massive number of  required correlation values leads to a substantial increase in computational load and a proportional decline in system efficiency \cite{DPE review}. To address this, employing a three-level (coarse, medium, fine) search grid can improve efficiency to some extent \cite{Collective detection}. Moreover, to reduce the risk of missing the CAF peak under a coarse step size, the correlation values around each grid point should be averaged \cite{Average CAF}. Another way to alleviate this dilemma is to utilize prior information from external sensors such as cellular network \cite{DPE GSM init} or INS \cite{DPE INS}, thereby refining the grid search parameters and improving the system efficiency. Unfortunately, even with multi-level or refined grid search parameters, the extensive computation of correlation values still relies on high-performance computing solutions, such as parallel processing \cite{Parallelized_DPE}. Consequently, some researches have proposed lightweight DPE framework using grid search \cite{Pseudorange DPE} or iterative optimization \cite{DBO DPE} at the pseudorange level. However, such approaches entail an inevitable trade-off: the simplification sacrifices the information in correlation domain, consequently lowering the theoretical performance upper bound of the system.

For improvement, the optimization instead of grid search have been applied in DPE, e.g.,  space alternating generalized expectation maximization (SAGE) \cite{SAGE}, accelerated random search (ARS) \cite{ARS_DPE}. 
Basically, these methods preliminary leverage individual correlation values for iterative optimization, still under the CAF maximization criterion. The convergency and accuracy of the iteration lack of theoretical guidance and the performance of such PVT estimation can not be guarantee. Inspired by the optimization concept, we found that through the proper characterization of the correlation values, i.e., spatial coherence and spatial gradient, it is possible to utilize the sparsely sampled spatial points to acquire necessary information for PVT estimation, thereby eliminating the dependency on the CAF peak. Therefore, we re-designed the optimization criterion from original CAF maximization to the CAF measurement residual minimization to jointly excavate PVT information contained within different spatial sampling points. The theoretical spatial sampled framework for DPE (SS-DPE) is proposed, as shown in Fig.\ref{fig:grid_opt_demo}(b), only sparse points are required in SS-DPE for PVT estimation, the computation load and system efficiency can be significantly improved.

\begin{figure*}[ht]
    \centering
    \includegraphics[width=0.8\linewidth]{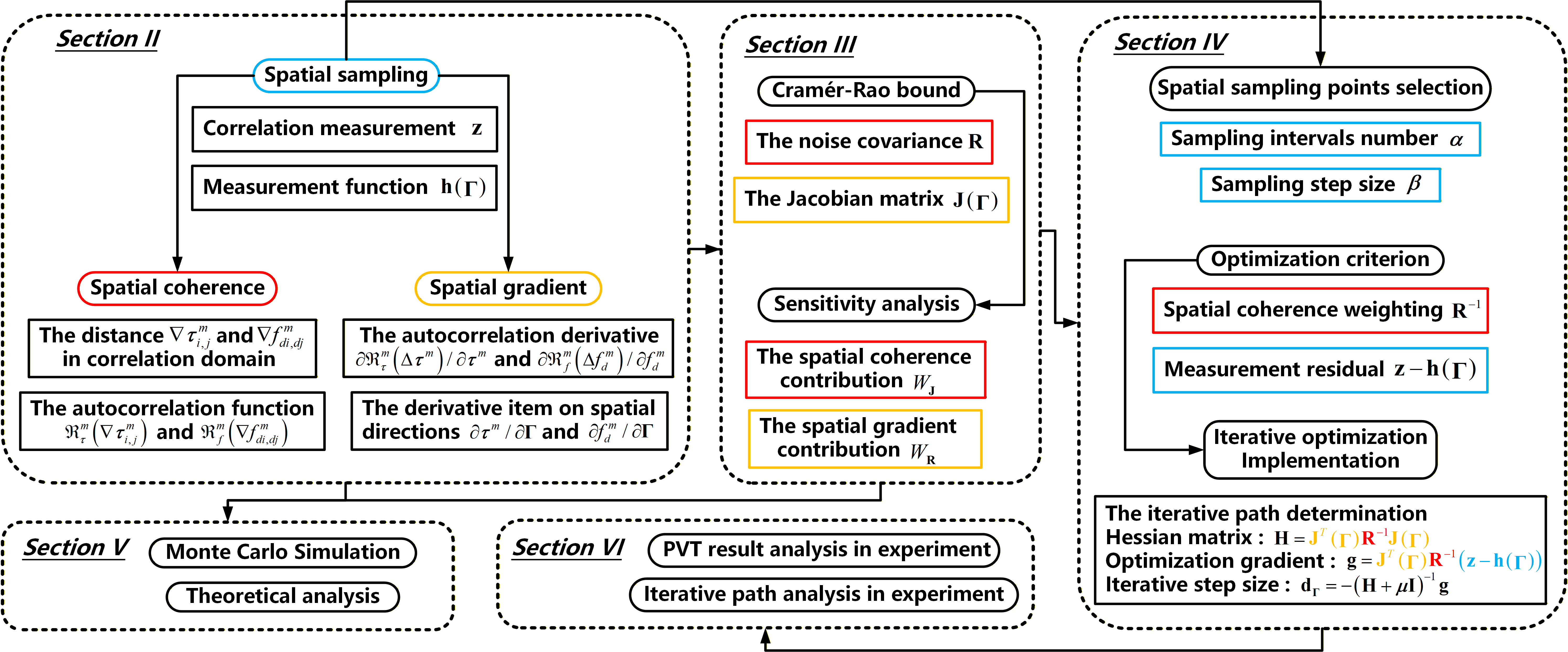}
    \caption{The theoretical framework of the proposed SS-DPE, where section II presents the correlation model of sparsely sampled points; section III derives the Cramér-Rao Bound; section IV proposes the optimization procedure; sections V and VI analyze the Monte Carlo simulation and experiment results.}
    \label{fig:theoretical_framework}
\end{figure*}

According to the above methodology, the total theoretical framework of the paper is shown as Fig. \ref{fig:theoretical_framework}. In Section II, we establish the measurement model that connects the correlation values with the PVT states to be estimated. Furthermore, the spatial coherence and gradient of these correlation values are accurately characterized, which in turn serve to quantify the diversity and density of PVT information, respectively. In Section III, the theoretical Cramér-Rao Bound (CRB) for PVT estimation using these correlation values are derived. The CRB is shown to depend entirely on the spatial coherence and spatial gradient, with their relative weighting further analyzed and quantified. The final iterative optimization framework for SS-DPE is implemented in Section IV. This framework comprises: (1) selecting spatial sampling points guided by the CRB and desired system characteristics; (2) formulating the optimization criterion utilizing all correlation measurements and their spatial properties; (3) applying gradient optimizers such as Levenberg-Marquardt (L-M) \cite{L-M optimization} to achieve PVT estimation. Subsequent Monte Carlo simulations in Section V validate the effectiveness of the correlation measurement model including its spatial properties, the Cramer-Rao bound derivation, and the iterative optimization framework for SS-DPE. Urban vehicle field experiments in Section VI further confirm the practical feasibility of the proposed SS-DPE framework.

The contributions of this paper can be summarized as follows: (1) the proposal of correlation measurement model for spatial sampling points, which employs spatial coherence and spatial gradient to characterize the diversity and density of the contained PVT information, respectively; (2) the derivation of the theoretical CRB for PVT estimation and its relationship with spatial properties, which is then applied to guide the selection of spatial sampling points; (3) the achievement of PVT parameter estimation under sparsely sampled points by fully utilizing all correlation measurements and their spatial properties, which significantly improves information utilization efficiency and reduces computation load.

\section{The correlation measurement model of spatial sampling points in DPE receiver}

In SS-DPE framework, the correlation value at each spatial sampling point serves as a direct measurement. These values are calculated from the raw signal data with the local replica. To establish their measurement model, we first revisit the raw intermediate frequency (IF) signal model, then derive the spatial coherence and gradient across different sampling points to fully characterize the correlation measurements.

\subsection{The linear transformation from raw signal samples to correlation measurementss}

In GNSS receivers, the continuous IF signals received from all visible satellites can be modeled as \cite{GNSS signal}
\begin{equation}
s\left( t \right) = \sum\limits_{m = 1}^M {{A^m}{C^m}\left( {t - {\tau ^m}/{f_c}} \right)\cos \left( {2\pi f_d^mt + {\varphi ^m}} \right) + \eta \left( t \right)}
\label{eq:IF_signal_t}
\end{equation}
where the superscript ‘$m$’ represents the corresponding parameters of the $m$-th satellite. The items $A^m$, $C^m\left(\cdot\right)$, and ${\eta \left( t \right)}$ represent the signal amplitude, pseudo-random noise (PRN) code, and noise item for received signals, respectively. ${f_c}$ is the PRN code frequency, and ${\varphi^m}$ is the carrier phase which can be regarded as an unknown term independent of PVT parameters in standard DPE receivers. ${\tau^m}$ and ${f_d^m}$ are code phase and carrier Doppler of each visible satellite respectively, which can be further expressed as \cite{Geometric DPE}
\begin{equation}
{\tau ^m} = \frac{{{f_c}}}{c}\left( {\left\| {{{\bf{r}}^m} - {\bf{r}}} \right\| + c\delta t - c\delta {t^m}} \right)
\label{eq:Tao}
\end{equation}
\begin{equation}
f_d^m =  - \frac{{{f_L}}}{c}\left[ {\frac{{{{\left( {{{\bf{r}}^m} - {\bf{r}}} \right)}^T}}}{{\left\| {{{\bf{r}}^m} - {\bf{r}}} \right\|}}\left( {{{\bf{v}}^m} - {\bf{v}}} \right) + c\delta \dot t - c\delta {{\dot t}^m}} \right]
\label{eq:Doppler}
\end{equation}
where ${\bf{r}} = {\left[ {\begin{array}{*{20}{c}}{{r_x}}&{{r_y}}&{{r_z}}\end{array}} \right]^T}$ and ${\bf{v}} = {\left[ {\begin{array}{*{20}{c}}{{v_x}}&{{v_y}}&{{v_z}}\end{array}} \right]^T}$ are the position and velocity of the receiver. $\delta t$ is the clock bias, and $\delta{\dot t}$ is the clock drift. $c$ is the speed of light, and ${f_L}$ is the carrier frequency.

The continuous IF signal expression of (\ref{eq:IF_signal_t}) is subsequently digitalized. Neglecting the quantization impact, the discretized IF signal model is shown as
\begin{equation}
s\left[ n \right] = \sum\limits_{m = 1}^M {{A^m}{C^m}\left( {n{T_s} - {\tau ^m}/{f_c}} \right)\cos \left( {2\pi f_d^mn{T_s} + {\varphi ^m}} \right)} + \eta \left[ n \right]
\label{eq:IF_signal_n}
\end{equation}
where the notation $s\left[ n \right] \buildrel\textstyle.\over= s\left( {n{T_s}} \right)$ indicates the discretized IF signal, obtained by sampling the continuous signal $s\left( t \right)$ with a sampling frequency ${f_s} = 1/{T_s}$. Considering all discretized points within a coherent integration time $T_{c}$, the length of the discrete sequence $s\left[ n \right]$ is $N = {T_c}{f_s}$. The Gaussian white noise sequence $\eta \left[ n \right]$ and amplitude ${A^m}$ of the signal satisfy the following expressions
\begin{equation}
E\left( {\eta \left[ n \right]{\eta ^T}\left[ n \right]} \right) = {\sigma ^2} = {N_0}{f_s}/2
\label{eq:Signal_noise}
\end{equation}
\begin{equation}
{A^m} = \sigma \sqrt {2C/N_0^m/{f_s}} 
\label{eq:Signal_amplitude}
\end{equation}
where $C/N^m_0$ is this the carrier-to-noise ratio which indicates the signal quality for each satellite channel.

Considering total number $N$ of IF sampling points within a coherent integration time $T_{c}$ , they can be listed as:
\begin{equation}
{\bf{s}} = {\left[ {\begin{array}{*{20}{c}}
{s\left[ 1 \right]}&{s\left[ 2 \right]}& \cdots &{s\left[ N \right]}
\end{array}} \right]^T}
\label{eq:IF_measurement_all}
\end{equation}
The sequence $\bf s$ contains the necessary information available for PVT estimation. 
In DPE receivers, the PVT parameters embedded in (\ref{eq:IF_signal_n}) and (\ref{eq:IF_measurement_all}) are direct values to be estimated, defined as \cite{DPE definition}
\begin{equation}
{\bf{\Gamma }} = {\left[ {\begin{array}{*{20}{c}}
{{{\bf{r}}^T}}&{\delta t}&{{{\bf{v}}^T}}&{\delta \dot t}
\end{array}} \right]^T}
\label{eq:PVT_define}
\end{equation}
The PVT parameters ${\bf{\Gamma }}$ are connected to the IF sampling points $s\left[ n \right]$ through code phase ${\tau ^m}$ in (\ref{eq:Tao}) and carrier Doppler ${f_d^m}$ in (\ref{eq:Doppler}). In situations where true PVT information is not available for implementation, we can preselect a series of spatial sampling points ${{\bf{\tilde \Gamma }}_1},{{\bf{\tilde \Gamma }}_2}, \cdots ,{{\bf{\tilde \Gamma }}_K}$ to reconstruct the local signals. For an arbitrary spatial sampling point $\bf{\tilde \Gamma }$, taking the in-phase (I) branch local signal as an example, the corresponding local signal can be expressed as
\begin{equation}
\tilde s_I^m\left[ {n,{\bf{\tilde \Gamma }}} \right] = {C^m}\left( {n{T_s} - {{\tilde \tau }^m}/{f_c}} \right)\cos \left( {2\pi \tilde f_d^mn{T_s}} \right)
\label{eq:Local_I_define}
\end{equation}
where the superscript ‘$(\tilde \cdot)$’ represents the parameters corresponds to spatial sampling point $\bf{\tilde \Gamma }$. The sampled code phase ${\tilde \tau }^m$ and carrier Doppler $\tilde f_d^m$ can be obtained by substitute spatial sampling point $\bf{\tilde \Gamma }$ into (\ref{eq:Tao}) and (\ref{eq:Doppler}). Then the correlation value corresponding to the selected spatial sampling point $\bf{\tilde \Gamma }$ are further calculated as
\begin{equation}
\Re _I^m\left( {{\bf{\tilde \Gamma }}} \right) = \frac{1}{N}\sum\limits_{n = 1}^N {s\left[ n \right]\tilde s_I^m\left[ {n,{\bf{\tilde \Gamma }}} \right]}
\label{eq:Correlation_I}
\end{equation}
It can be observed from (\ref{eq:Correlation_I}) that the correlation value ${\Re _I^m\left( {{\bf{\tilde \Gamma }}} \right)}$ is a linear combination of the IF sampling points $s\left[ n \right]$. Since all IF sampling points contain the complete PVT information embedded in GNSS signals, the correlation values obtained through linear transformation of IF sampling points likewise preserve this PVT information. Therefore, in principle, correlation values can also serve as the measurement for the estimator of PVT parameters.

In order to use the correlation values as measurements, it is necessary to establish their relationship with the PVT states to be estimated. Substitute (\ref{eq:IF_signal_n}) and (\ref{eq:Local_I_define}) into (\ref{eq:Correlation_I}), then the measurement equation of correlation values are derived as
\begin{equation}
\begin{aligned}
&\Re _I^m\left( {{\bf{\tilde \Gamma }}} \right)\\
 =& \frac{1}{N}\sum\limits_{n = 1}^N \left\{ \tilde s_I^m\left[ {n,{\bf{\tilde \Gamma }}} \right] \sum\limits_{m = 1}^M \left[ {A^m}{C^m}\left( {n{T_s} - {\tau ^m}/{f_c}} \right) \right. \right. \\
 &\quad \left. \left. \times \cos \left( {2\pi f_d^m n{T_s} + {\varphi ^m}} \right) \right] \right\}
  + \frac{1}{N}\sum\limits_{n = 1}^N \tilde s_I^m\left[ {n,{\bf{\tilde \Gamma }}} \right]\eta \left[ n \right] \\
 \approx& \frac{{{A^m}}}{N}\sum\limits_{n = 1}^N \left\{ {C^m}\left( {n{T_s} - {{\tilde \tau }^m}/{f_c}} \right){C^m}\left( {n{T_s} - {\tau ^m}/{f_c}} \right) \right. \cos \left( {2\pi \tilde f_d^m n{T_s}} \right)\\
 &\quad \left. \times  \cos \left( {2\pi f_d^m n{T_s} + {\varphi ^m}} \right) \right\}+ \frac{1}{N}\sum\limits_{n = 1}^N \tilde s_I^m\left[ {n,{\bf{\tilde \Gamma }}} \right]\eta \left[ n \right] \\
 \approx& \frac{1}{2}{A^m}\Re _\tau ^m\left( {\Delta {\tau ^m}} \right)\Re _f^m\left( {\Delta f_d^m} \right)\cos \left( {{\varphi ^m}} \right)
  + \frac{1}{N}\sum\limits_{n = 1}^N \tilde s_I^m\left[ {n,{\bf{\tilde \Gamma }}} \right]\eta \left[ n \right] \\
 \buildrel\textstyle.\over=& h_I^m\left( {{\bf{\tilde \Gamma }},{\bf{\Gamma }}} \right) + \nu _I^m\left( {{\bf{\tilde \Gamma }}} \right)
\end{aligned}
\label{eq:I_model}
\end{equation}
where $\Delta {\tau ^m} = {\tau ^m} - {\tilde \tau ^m}$ and $\Delta f_d^m = f_d^m - \tilde f_d^m$ represent the code phase error and Doppler error between sampled values and true values. The derivation process of (\ref{eq:I_model}) utilizes the auto-correlation and cross-correlation properties of the PRN code $C^m\left(\cdot\right)$. The specific expressions for correlation functions $\Re _\tau ^m\left( \cdot \right)$ and $\Re _f^m\left( \cdot \right)$ are given as \cite{Parallelized_DPE}
\begin{equation}
\Re _\tau ^m\left( {\Delta {\tau ^m}} \right) \approx \max \left( {0,1 - \left| {\Delta {\tau ^m}} \right|} \right)
\label{eq:CAF_tao}
\end{equation}
\begin{equation}
\Re _f^m\left( {\Delta f_d^m} \right) = \frac{{{\rm{sin}}\left( {\pi \Delta f_d^m{T_c}} \right)}}{{\pi \Delta f_d^m{T_c}}} \buildrel\textstyle.\over= {\rm{sinc}}\left( {\Delta f_d^m{T_c}} \right)
\label{eq:CAF_Doppler}
\end{equation}
Thus, we determine the specific nonlinear function ${h^m_I}\left( {{\bf{\tilde \Gamma }},{\bf{\Gamma }}} \right)$ between the correlation values ${\Re _I^m\left( {{\bf{\tilde \Gamma }}} \right)}$ and the PVT states ${\bf \Gamma }$. In other words, the measurement function ${h^m_I}\left( {{\bf{\tilde \Gamma }},{\bf{\Gamma }}} \right)$ reveals the PVT information contained in the correlation measurement ${\Re _I^m\left( {{\bf{\tilde \Gamma }}} \right)}$ corresponding to specific spatial sampling points ${\bf{\tilde \Gamma }}$. This constitutes the theoretical foundation for utilizing correlation measurements to estimate PVT parameters. 

For all available spatial sampling points ${{\bf{\tilde \Gamma }}_1},{{\bf{\tilde \Gamma }}_2}, \cdots ,{{\bf{\tilde \Gamma }}_K}$ and satellite channel $m = 1,2, \cdots ,M$, the total correlation measurements ${\bf{z}}$ is expressed in terms of all spatial sampling points across all satellite channels.
\begin{equation}
{{\bf{z}}^m} = {\left[ {\begin{array}{*{20}{c}}
{\Re _I^m\left( {{{{\bf{\tilde \Gamma }}}_1}} \right)}&{\Re _I^m\left( {{{{\bf{\tilde \Gamma }}}_2}} \right)}& \cdots &{\Re _I^m\left( {{{{\bf{\tilde \Gamma }}}_K}} \right)}
\end{array}} \right]^T}
\label{eq:Measurement_single}
\end{equation}
\begin{equation}
{\bf{z}} = {\left[ {\begin{array}{*{20}{c}}
{{{\left( {{{\bf{z}}^1}} \right)}^T}}&{{{\left( {{{\bf{z}}^2}} \right)}^T}}& \cdots &{{{\left( {{{\bf{z}}^M}} \right)}^T}}
\end{array}} \right]^T}
\label{eq:Measurement_all}
\end{equation}
Similarly, we define the total measurement function vector $\mathbf{h}(\boldsymbol{\Gamma})$ from single measurement function ${h^m_I}\left( {{\bf{\tilde \Gamma }},{\bf{\Gamma }}} \right)$ and the total measurement noise vector $\boldsymbol{\nu}$ from single measurement noise $\nu _I^m\left( {{\bf{\tilde \Gamma }}} \right)$. Consequently, all PVT information originally in the IF samples $\mathbf{s}$ is now fully encapsulated in the correlation measurements $\mathbf{z}$ and their model $\mathbf{h}(\boldsymbol{\Gamma})$, with the original IF noise $\eta[n]$ transformed into the correlation-domain noise $\boldsymbol{\nu}$.

\subsection{The spatial coherence for spatial sampling points}

When a series of spatial sampling points ${{\bf{\tilde \Gamma }}_1},{{\bf{\tilde \Gamma }}_2}, \cdots ,{{\bf{\tilde \Gamma }}_K}$ are selected, the spatial coherence among their corresponding correlation values $\Re _I^m\left( {{{{\bf{\tilde \Gamma }}}_1}} \right),\Re _I^m\left( {{{{\bf{\tilde \Gamma }}}_2}} \right), \cdots ,\Re _I^m\left( {{{{\bf{\tilde \Gamma }}}_K}} \right)$ will also become apparent. The physical meaning of spatial coherence is the diversity in PVT information across different spatial sampling points, which is mathematically characterized by the covariance of the correlation measurement noise.

Due to the linear transformation relationship from IF sampling point measurements $s\left[ n \right]$ to correlation measurements ${\Re _I^m\left( {{\bf{\tilde \Gamma }}} \right)}$ shown in (\ref{eq:Correlation_I}), the noise $\eta \left[ n \right]$ is also transformed to $\nu _I^m\left( {{\bf{\tilde \Gamma }}} \right)$ accordingly as shown in (\ref{eq:I_model}). 
\begin{equation}
\nu _I^m\left( {{\bf{\tilde \Gamma }}} \right) = \frac{1}{N}\sum\limits_{n = 1}^N {\tilde s_I^m\left[ {n,{\bf{\tilde \Gamma }}} \right]\eta \left[ n \right]} 
\label{eq:noise_model}
\end{equation}
As a result, the spatial coherence among different spatial sampling points is reflected in the noise cross-covariance induced by the transformation.

Assuming two spatial sampling points ${\bf{\tilde \Gamma}}_i$ and ${\bf{\tilde \Gamma}}_j$, under the same $m$-th satellite, their corresponding measurement noises for the coherent correlation values $\Re _I^m\left( {{{{\bf{\tilde \Gamma }}}_i}} \right)$ and $\Re _I^m\left( {{{{\bf{\tilde \Gamma }}}_j}} \right)$ are respectively given as ${\nu_I^m}\left( {{{{\bf{\tilde \Gamma }}}_i}} \right)$ and ${\nu_I^m}\left( {{{{\bf{\tilde \Gamma }}}_j}} \right)$. Using the expression of $\nu _I^m\left( {{\bf{\tilde \Gamma }}} \right)$ shown in (\ref{eq:noise_model}) as well as the local signal expression $\tilde s_I^m\left[ {n,{\bf{\tilde \Gamma }}} \right]$ shown in (\ref{eq:Local_I_define}), the cross-covariance can be derived as
\begin{equation}
\begin{aligned}
&E\left[ {\nu _I^m\left( {{{{\bf{\tilde \Gamma }}}_i}} \right)\nu _I^m\left( {{{{\bf{\tilde \Gamma }}}_j}} \right)} \right]\\
 =& E\left[ {\frac{1}{N}\sum\limits_{n = 1}^N {\left( {\tilde s_I^m\left[ {n,{{{\bf{\tilde \Gamma }}}_i}} \right]\eta \left[ n \right]} \right)}  \cdot \frac{1}{N}\sum\limits_{n = 1}^N {\left( {\tilde s_I^m\left[ {n,{{{\bf{\tilde \Gamma }}}_j}} \right]\eta \left[ n \right]} \right)} } \right]\\
 =& \frac{1}{{{N^2}}}\sum\limits_{n = 1}^N {\tilde s_I^m\left[ {n,{{{\bf{\tilde \Gamma }}}_i}} \right]\tilde s_I^m\left[ {n,{{{\bf{\tilde \Gamma }}}_j}} \right]} E\left( {\eta \left[ n \right]{\eta ^T}\left[ n \right]} \right)\\
 =& \frac{{{N_0}{f_s}}}{{2{N^2}}}\sum\limits_{n = 1}^N {\left[ {{C^m}\left( {n{T_s} - \tilde \tau _i^m/{f_c}} \right){C^m}\left( {n{T_s} - \tilde \tau _j^m/{f_c}} \right)} \right.} \\
 &\left. { * \cos \left( {2\pi \tilde f_{di}^mn{T_s}} \right)\cos \left( {2\pi \tilde f_{dj}^mn{T_s}} \right)} \right]\\
 \approx& \frac{{{N_0}}}{{4{T_c}}}\Re _\tau ^m\left( {\nabla \tau _{i,j}^m} \right)\Re _f^m\left( {\nabla f_{di,dj}^m} \right)
\end{aligned}
\label{eq:variance_ij}
\end{equation}
The derivation process of (\ref{eq:variance_ij}) utilizes the time-uncorrelated property of the white noise sequence $\eta \left[ n \right]$, shown as $E\left( {\eta \left[ {{n_1}} \right]{\eta ^T}\left[ {{n_2}} \right]} \right) = 0\left( {{n_1} \ne {n_2}} \right)$. The summation term in (\ref{eq:variance_ij}) can be approximated as an integral and described using the autocorrelation functions $\Re _\tau ^m\left( \cdot \right)$ of (\ref{eq:CAF_tao}) and $\Re _f^m\left( \cdot \right)$ of (\ref{eq:CAF_Doppler}). $\nabla \tau _{i,j}^m = \tilde \tau _i^m - \tilde \tau _j^m$ and $\nabla f_{di,dj}^m = \tilde f_{di}^m - \tilde f_{dj}^m$ represent the distance in equivalent code phase and Doppler between two spatial sampling points ${\bf{\tilde \Gamma}}_i$ and ${\bf{\tilde \Gamma}}_j$. In particular, for each correlation value $\Re _I^m\left( {{{{\bf{\tilde \Gamma }}}_i}} \right)$ corresponding to spatial sampling point ${{{\bf{\tilde \Gamma }}}_i}$, the noise variance $E\left[ {\nu _I^m{{\left( {{{{\bf{\tilde \Gamma }}}_i}} \right)}^2}} \right]$ holds fixed value of ${{N_0}/\left( {4{T_c}} \right)}$.

Next, we derive the cross-covariance of the correlation values between different satellites. Assuming two spatial sampling points ${\bf{\tilde \Gamma}}_i$ and ${\bf{\tilde \Gamma}}_j$, under different satellites $m_1$ and $m_2$ respectively (${m_1} \ne {m_2}$), the cross-covariance for measurement noises ${\nu_I^{m_1}}\left( {{{{\bf{\tilde \Gamma }}}_i}} \right)$ and ${\nu_I^{m_2}}\left( {{{{\bf{\tilde \Gamma }}}_j}} \right)$ can be obtained as
\begin{equation}
\begin{aligned}
&E\left[ {\nu _I^{{m_1}}\left( {{{{\bf{\tilde \Gamma }}}_i}} \right)\nu _I^{{m_2}}\left( {{{{\bf{\tilde \Gamma }}}_j}} \right)} \right]\\
 \approx& \frac{{{N_0}}}{{4{T_c}}}\Re _\tau ^{{m_1},{m_2}}\left( {\nabla \tau _{i,j}^{{m_1},{m_2}}} \right)\Re _f^{{m_1},{m_2}}\left( {\nabla f_{di,dj}^{{m_1},{m_2}}} \right)
\end{aligned}
\label{eq:variance_m1m2}
\end{equation}
where $\nabla \tau _{i,j}^{{m_1},{m_2}} = \tilde \tau _i^{{m_1}} - \tilde \tau _j^{{m_2}}$ and $\nabla f_{di,dj}^{{m_1},{m_2}} = \tilde f_{di}^{{m_1}} - \tilde f_{dj}^{{m_2}}$ are the equivalent code phase distance across different satellites. The expression of the carrier cross-correlation function $\Re _f^{{m_1},{m_2}}\left( \cdot \right)$ is the same as (\ref{eq:CAF_Doppler}), while the code cross-correlation function $\Re _\tau ^{{m_1},{m_2}}\left( \cdot \right) \approx 0$ due to the cross-correlation properties of the PRN code. Therefore, for the correlation measurements of different satellites, their cross-covariance satisfies $E\left[ {\nu _I^{m_1}\left( {{{{\bf{\tilde \Gamma }}}_i}} \right)\nu _I^{m_2}\left( {{{{\bf{\tilde \Gamma }}}_j}} \right)} \right] \approx 0$, which implies a negligible spatial coherence among correlation values from different satellites.

In summary, the measurement noise covariance matrix $\bf R$ can be approximately expressed in a block-diagonal form, where each block ${\bf R}^m$ quantifies the spatial coherence among the spatial sampling points.
\begin{equation}
{{\bf{R}}^m} = E\left[ {{{\bm{\nu }}^m}{{\left( {{{\bm{\nu }}^m}} \right)}^T}} \right] = {\left[ {\frac{{{N_0}}}{{4{T_c}}}\Re _\tau ^m\left( {\nabla \tau _{i,j}^m} \right)\Re _f^m\left( {\nabla f_{di,dj}^m} \right)} \right]_{K \times K}}
\label{eq:R_single}
\end{equation}
\begin{equation}
{\bf{R}} = E\left[ {{\bm{\nu}}{{\bm{\nu}}^T}} \right] = blkdiag\left( {\left[ {\begin{array}{*{20}{c}}
{{{\bf{R}}^1}}&{{{\bf{R}}^2}}& \cdots &{{{\bf{R}}^M}}
\end{array}} \right]} \right)
\label{eq:R_all}
\end{equation}
where ${\left[ {f\left( {i,j} \right)} \right]_{K \times K}}$ denotes a $K\times K$ matrix whose $(i,j)$-th element is given by $f(i,j)$. The spatial coherence between spatial sampling points exists within the same satellite channel and can be precisely characterized by the equivalent code phase distance $\nabla \tau _{i,j}^m$ and Doppler distance $\nabla f_{di,dj}^m$ between these spatial sampling points. A smaller code phase distance $\nabla \tau _{i,j}^m$ and Doppler distance $\nabla f_{di,dj}^m$ imply greater spatial coherence, meaning a less diversity of PVT information between the two spatial points.

\subsection{The spatial gradient for spatial sampling points}

The spatial coherence quantifies the diversity of PVT information contained in the correlation values across different sampling points. A complementary quantity, the spatial gradient, is thus introduced to characterize the PVT information density within a single point. Mathematically, the spatial gradient can be described by the Jacobian matrix $\partial h_I^m\left( {{\bf{\tilde \Gamma }},{\bf{\Gamma }}} \right)/\partial {\bf{\Gamma }} $ of the correlation measurement function ${h^m_I}\left( {{\bf{\tilde \Gamma }},{\bf{\Gamma }}} \right)$.

Considering the measurement equation of a certain correlation value ${\Re _I^m\left( {{\bf{\tilde \Gamma }}} \right)}$, the measurement function for it has been derived in (\ref{eq:I_model}), given as 
\begin{equation}
h_I^m\left( {{\bf{\tilde \Gamma }},{\bf{\Gamma }}} \right) = \frac{1}{2}{A^m}\Re _\tau ^m\left( {\Delta {\tau ^m}} \right)\Re _f^m\left( {\Delta f_d^m} \right)\cos \left( { \varphi ^m} \right)
\label{eq:Measurement_equation}
\end{equation}
Using the expressions of correlation functions $\Re _\tau ^m\left( \cdot \right)$ and $\Re _f^m\left( \cdot \right)$ in (\ref{eq:CAF_tao}) and (\ref{eq:CAF_Doppler}), the partial derivative of $h_I^m\left( {{\bf{\tilde \Gamma }},{\bf{\Gamma }}} \right)$ with respect to PVT parameters $\bf{\Gamma }$ is given as
\begin{equation}
\begin{split}
\frac{{\partial h_I^m\left( {{\bf{\tilde \Gamma }},{\bf{\Gamma }}} \right)}}{{\partial {\bf{\Gamma }}}}
&= \frac{{A^m}}{2}\cos \left( {{\varphi ^m}} \right) 
\left[ \frac{{\partial \Re _\tau ^m\left( {\Delta {\tau ^m}} \right)}}{{\partial {\tau ^m}}}
\frac{{\partial {\tau ^m}}}{{\partial {\bf{\Gamma }}}}
\Re _f^m\left( {\Delta f_d^m} \right) \right. \\
&\quad \left. + \frac{{\partial \Re _f^m\left( {\Delta f_d^m} \right)}}{{\partial f_d^m}}
\frac{{\partial f_d^m}}{{\partial {\bf{\Gamma }}}}
\Re _\tau ^m\left( {\Delta {\tau ^m}} \right) \right]
\end{split}
\label{eq:Jacobian_main}
\end{equation}
The partial derivative terms $\partial \Re_\tau^m (\Delta \tau^m) / \partial \tau^m$ and $\partial \Re _f^m\left( {\Delta f_d^m} \right)/\partial f_d^m$ are further derived as

\begin{equation}
\frac{{\partial \Re _\tau ^m\left( {\Delta {\tau ^m}} \right)}}{{\partial {\tau ^m}}} =  - rect\left( {\frac{{\Delta {\tau ^m}}}{2}} \right){\rm{sig}}n\left( {\Delta {\tau ^m}} \right)
\label{eq:Jacobian_tao}
\end{equation}
\begin{equation}
\frac{{\partial \Re _f^m\left( {\Delta f_d^m} \right)}}{{\partial f_d^m}} = \frac{{\left( {\pi \Delta f_d^m{T_c}} \right)\cos \left( {\pi \Delta f_d^m{T_c}} \right) - \sin \left( {\pi \Delta f_d^m{T_c}} \right)}}{{\pi {T_c}{{\left( {\Delta f_d^m} \right)}^2}}}
\label{eq:Jacobian_Doppler}
\end{equation}
where $sign\left(  \cdot  \right)$ denotes the signum function, and $rect\left(  \cdot  \right)$ denotes the rectangular window function.
Due to the property of the $\text{rect}(\cdot)$ function, $\partial h_I^m(\tilde{\boldsymbol{\Gamma}}, \boldsymbol{\Gamma}) / \partial \boldsymbol{\Gamma} = 0$ when $|\Delta \tau^m| > 1$. The unknown carrier phase parameter ${\varphi ^m}$ in (\ref{eq:Jacobian_main}) needs to be obtained through the $arctan$ type of phase discriminator, using the $I/Q$ channel correlation values
\begin{equation}
{\varphi ^m} = \arctan \left( {\Re _Q^m\left( {{\bf{\tilde \Gamma }}} \right)/\Re _I^m\left( {{\bf{\tilde \Gamma }}} \right)} \right)
\label{eq:IQ_arctan2}
\end{equation}
and the signal amplitude $A^m$ in (\ref{eq:Jacobian_main}) can be determined using (\ref{eq:Signal_noise}) and (\ref{eq:Signal_amplitude}). It can be seen that the remaining unknown items in (\ref{eq:Jacobian_main}) are the partial derivatives of the code phase ${\tau ^m}$ and carrier Doppler $f_d^m$ with respect to PVT parameters $\bf{\Gamma }$. Specifically, they refer to  $\partial {\tau ^m}/\partial {\bf{\Gamma }}$ and $\partial f_d^m/\partial {\bf{\Gamma }}$.

According to the expression of code phase ${\tau ^m}$ and carrier Doppler $f_d^m$ shown in (\ref{eq:Tao}) and (\ref{eq:Doppler}) and the definition of PVT parameters $\bf{\Gamma }$ shown in (\ref{eq:PVT_define}), the detailed partial derivative terms can be derived for  $\partial {\tau ^m}/\partial {\bf{\Gamma }}$ and $\partial f_d^m/\partial {\bf{\Gamma }}$ as shown in Table\ref{eq:Partial_derivative_items}, where ${{\bf{u}}^m} = \left( {{{\bf{r}}^m} - {\bf{r}}} \right)/\left\| {{{\bf{r}}^m} - {\bf{r}}} \right\|$ represent the LOS direction vector. 

\begin{table}[ht]
\centering
\caption{The detailed partial derivative items for $\partial {\tau ^m}/\partial {\bf{\Gamma }}$ and $\partial f_d^m/\partial {\bf{\Gamma }}$}
\begin{tabular}{|c|c|c|}
\hline
 & ${\partial {\tau ^m}}$ & ${\partial f_d^m}$ \\ \hline
$/{\partial {\bf{r}}}$& $ - \frac{{{f_c}}}{c}{\left( {{{\bf{u}}^m}} \right)^T}$ & $\frac{{{f_L}}}{c}{\left( {{{\bf{v}}^m} - {\bf{v}}} \right)^T}\frac{{{\bf{I}} - {{\bf{u}}^m}{{\left( {{{\bf{u}}^m}} \right)}^T}}}{{\left\| {{{\bf{r}}^m} - {\bf{r}}} \right\|}}$ \\ \hline
$/{\partial \delta t}$& ${f_c}$ & $0$ \\ \hline
$/{\partial {\bf{v}}}$& ${\bf{0}}$ & $\frac{{{f_L}}}{c}{\left( {{{\bf{u}}^m}} \right)^T}$ \\ \hline
$/{\partial \delta \dot t}$& $0$ & $- {f_L}$ \\ \hline
\end{tabular}
\label{eq:Partial_derivative_items}
\end{table}
Using the detailed partial derivative items in Table \ref{eq:Partial_derivative_items} as well as the main partial derivative expression (\ref{eq:Jacobian_main}), we can finally obtain the whole Jacobian matrix for all available correlation measurements shown as (\ref{eq:Jacobian_single}) and (\ref{eq:Jacobian_all})
\begin{equation}
{{\bf{J}}^m}\left( {\bf{\Gamma }} \right) = {\left[ {\begin{array}{*{20}{c}}
{{{\frac{{\partial h_I^m\left( {{{{\bf{\tilde \Gamma }}}_1},{\bf{\Gamma }}} \right)}}{{\partial {\bf{\Gamma }}}}}^T}}& \cdots &{{{\frac{{\partial h_I^m\left( {{{{\bf{\tilde \Gamma }}}_K},{\bf{\Gamma }}} \right)}}{{\partial {\bf{\Gamma }}}}}^T}}
\end{array}} \right]^T}
\label{eq:Jacobian_single}
\end{equation}
\begin{equation}
{\bf{J}}\left( {\bf{\Gamma }} \right) = {\left[ {\begin{array}{*{20}{c}}
{{{\bf{J}}^1}{{\left( {\bf{\Gamma }} \right)}^T}}&{{{\bf{J}}^2}{{\left( {\bf{\Gamma }} \right)}^T}}& \cdots &{{{\bf{J}}^M}{{\left( {\bf{\Gamma }} \right)}^T}}
\end{array}} \right]^T}
\label{eq:Jacobian_all}
\end{equation}

Equation (\ref{eq:Jacobian_main}) shows that the spatial gradient in the position-clock bias $\left[ {{\bf{r}},\delta t} \right]$ dimension is governed by two terms: $\partial \Re_\tau^m (\Delta \tau^m) / \partial \tau^m$ and $\partial \tau^m / \partial \boldsymbol{\Gamma}$. The first term captures the intrinsic gradient of the correlation function $\Re _\tau ^m\left( \cdot \right)$, while the second projects it onto the spatial coordinates of $[\mathbf{r}, \delta t]$. Physically, this means the PVT information density in $\Re_I^m(\tilde{\boldsymbol{\Gamma}})$ is first described by the gradient with respect to the code phase error $\Delta \tau^m$, and then spatially distributed via the geometric projection $\partial \tau^m / \partial \boldsymbol{\Gamma}$. The similar analysis is also applicable for velocity-clock drift $\left[ {{\bf{v}},\delta \dot t} \right]$.

\section{The Cramér-Rao Bound analysis for the spatial sampling-based DPE}

The correlation values from all spatial sampling points ${{\bf{\tilde \Gamma }}_1},{{\bf{\tilde \Gamma }}_2}, \cdots ,{{\bf{\tilde \Gamma }}_K}$ collectively form the information used for PVT estimation. The Cramér-Rao Bound represents the theoretical performance limit for PVT parameter estimation utilizing the spatial information in SS-DPE framework.

\subsection{The theoretical Cramér-Rao Bound for PVT estimation}

The correlation values corresponding to multiple spatial points contain the PVT information of the original IF signal, due to their linear transformation relationship. When using these correlation values to estimate PVT parameters, there exists a theoretical accuracy bound expressed as \cite{CRB define}
\begin{equation}
{\bf{P}}\left( {{\bf{\hat \Gamma }}} \right) \ge {\bf{P}}\left( {\bf{\Gamma }} \right) = {\bf{F}}{\left( {\bf{\Gamma }} \right)^{ - 1}}
\label{eq:CRB_define}
\end{equation}
where ${\bf{P}}\left( {{\bf{\hat \Gamma }}} \right)$ represents the covariance matrix for the estimation results. ${\bf{F}}\left( {\bf{\Gamma }} \right)$ is commonly referred to as the Fisher information matrix (FIM), whose inverse is the Cramér-Rao Bound matrix ${\bf{P}}\left( {\bf{\Gamma }} \right)$.

The measurement model for total correlation values can be written as
\begin{equation}
{\bf{z}} = {\bf{h}}\left( {\bf{\Gamma }} \right) + {\bm{\nu }}
\label{eq:Measurement_model_all}
\end{equation}
where ${\bf{z}}$ is the total correlation measurements defined in (\ref{eq:Measurement_all}), with total measurement function ${\bf{h}}\left( {\bf{\Gamma }} \right)$ and total measurement noise $\bm \nu$. The elements of FIM ${\bf{F}}\left( {\bf{\Gamma }} \right)$ are defined by
\begin{equation}
{\left[ {{\bf{F}}\left( {\bf{\Gamma }} \right)} \right]_{u,v}} = {\rm E}\left( {\frac{{{\partial ^2}\ln p\left( {{\bf{z}}|{\bf{\Gamma }}} \right)}}{{\partial {\Gamma _u}\partial {\Gamma _v}}}} \right)
\label{eq:FIM_define}
\end{equation}
where the symbol ${\left[  \cdot  \right]_{u,v}}$ represents the element in the $u$-th row and $v$-th column of the matrix, ${{\Gamma _u}}$ is the $u$-th element of PVT parameters ${\bf{\Gamma }}$. According  to the measurement model expression of (\ref{eq:Measurement_model_all}), the likelihood probability density function ${p\left( {{\bf{z}}|{\bf{\Gamma }}} \right)}$ follows a multivariate Gaussian distribution, which can be expressed as
\begin{equation}
p\left( {{\bf{z}}|{\bf{\Gamma }}} \right) = \frac{1}{{{{\left( {2\pi } \right)}^{\frac{{KM}}{2}}}{{\left| {\bf{R}} \right|}^{\frac{1}{2}}}}}\exp \left[ { - \frac{1}{2}{{\left( {{\bf{z}} - {\bf{h}}\left( {\bf{\Gamma }} \right)} \right)}^T}{{\bf{R}}^{ - 1}}\left( {{\bf{z}} - {\bf{h}}\left( {\bf{\Gamma }} \right)} \right)} \right]
\label{eq:likelihood_PDF}
\end{equation}
Substitute (\ref{eq:likelihood_PDF}) into (\ref{eq:FIM_define}), the elements of FIM expression ${\bf{F}}\left( {\bf{\Gamma }} \right)$ can be derived as
\begin{equation}
{\left[ {{\bf{F}}\left( {\bf{\Gamma }} \right)} \right]_{u,v}} = {\frac{{\partial {\bf{h}}\left( {\bf{\Gamma }} \right)}}{{\partial {\Gamma _u}}}^T}{{\bf{R}}^{ - 1}}\frac{{\partial {\bf{h}}\left( {\bf{\Gamma }} \right)}}{{\partial {\Gamma _v}}} + \frac{1}{2}tr\left( {{{\bf{R}}^{ - 1}}\frac{{\partial {\bf{R}}}}{{\partial {\Gamma _u}}}{{\bf{R}}^{ - 1}}\frac{{\partial {\bf{R}}}}{{\partial {\Gamma _v}}}} \right)
\label{eq:FIM_element}
\end{equation}
where $tr$ represents the trace of the matrix. According to the expression of (\ref{eq:R_single})-(\ref{eq:R_all}) derived above, The noise covariance matrix ${\bf{R}}$ depends only on the selection of spatial sampling points  ${{\bf{\tilde \Gamma }}_1},{{\bf{\tilde \Gamma }}_2}, \cdots ,{{\bf{\tilde \Gamma }}_K}$ and the signal noise parameter $N_0$ , and is independent of the true PVT parameters ${\bf{\Gamma }}$. Hence we have $\partial {\bf{R}}/\partial {\Gamma _u} = {\bf{0}}$, and only the first item in (\ref{eq:FIM_element}) is useful. By integrating the expressions (\ref{eq:FIM_element}) into matrix form, we can finally obtain the expression of theoretical Cramér-Rao Bound
\begin{equation}
{\bf{P}}\left( {{\bf{\hat \Gamma }}} \right) \ge {\bf{P}}\left( {\bf{\Gamma }} \right) = {\bf{F}}{\left( {\bf{\Gamma }} \right)^{ - 1}} = {\left[ {{{\bf{J}}^T}\left( {\bf{\Gamma }} \right){{\bf{R}}^{ - 1}}{{\bf{J}}^T}\left( {\bf{\Gamma }} \right)} \right]^{ - 1}}
\label{eq:CRB_bound}
\end{equation}
where ${\bf{J}}\left( {\bf{\Gamma }} \right)$ represents the Jacobian matrix for correlation measurements. Expression (\ref{eq:CRB_bound}) reveals that the Cramér-Rao Bound $\mathbf{P}(\boldsymbol{\Gamma})$ for PVT estimation is determined jointly by the noise covariance matrix $\mathbf{R}$ and the Jacobian matrix $\mathbf{J}(\boldsymbol{\Gamma})$, as analyzed earlier. This shows that the theoretical performance limit depends on both the PVT information density at individual sampling points (governed by $\mathbf{J}$) and the PVT information diversity among different points (captured by $\mathbf{R}$).

\subsection{The sensitivity analysis for Cramér-Rao Bound}

Given that the Cramér-Rao bound $\mathbf{P}(\hat{\boldsymbol{\Gamma}})$ depends on both $\mathbf{R}$ and $\mathbf{J}(\boldsymbol{\Gamma})$, a key question arises: what are their relative contributions, and how can they be quantitatively assessed and compared?

In practice, we typically focus only on the diagonal elements of the Cramér-Rao bound, which describe the estimation performance of PVT parameters such as position-clock bias and velocity-clock drift. The trace of the Cramér-Rao bound matrix can be used to provide an overall description. Therefore, the contributions of ${\bf{R}}$ and ${\bf{J}}\left( {\bf{\Gamma }} \right)$ can be described by the partial derivatives of the trace for Cramér-Rao matrix with respect to ${\bf{R}}$ and ${\bf{J}}\left( {\bf{\Gamma }} \right)$, which are shown as $\partial tr\left( {{\bf{P}}\left( {\bf{\Gamma }} \right)} \right)/\partial {\bf{J}}\left( {\bf{\Gamma }} \right)$ and $\partial tr\left( {{\bf{P}}\left( {\bf{\Gamma }} \right)} \right)/\partial {\bf{R}}$. The specific forms of these two expressions are obtained by taking the partial derivatives of (\ref{eq:CRB_bound}), utilizing some properties of the matrix trace and matrix derivatives.
\begin{equation}
\frac{{\partial tr\left( {{\bf{P}}\left( {\bf{\Gamma }} \right)} \right)}}{{\partial {\bf{J}}\left( {\bf{\Gamma }} \right)}} =  - 2{{\bf{R}}^{ - 1}}{\bf{J}}\left( {\bf{\Gamma }} \right){\bf{P}}\left( {\bf{\Gamma }} \right){{\bf{P}}^T}\left( {\bf{\Gamma }} \right)
\label{eq:sensitivity_J}
\end{equation}
\begin{equation}
\frac{{\partial tr\left( {{\bf{P}}\left( {\bf{\Gamma }} \right)} \right)}}{{\partial {\bf{R}}}} = {{\bf{R}}^{ - 1}}{\bf{J}}\left( {\bf{\Gamma }} \right){\bf{P}}\left( {\bf{\Gamma }} \right){{\bf{P}}^T}\left( {\bf{\Gamma }} \right){{\bf{J}}^T}\left( {\bf{\Gamma }} \right){{\bf{R}}^{ - 1}}
\label{eq:sensitivity_R}
\end{equation}
The partial derivative expression of $\partial tr\left( {{\bf{P}}\left( {\bf{\Gamma }} \right)} \right)/\partial {\bf{J}}\left( {\bf{\Gamma }} \right)$ and $\partial tr\left( {{\bf{P}}\left( {\bf{\Gamma }} \right)} \right)/\partial {\bf{R}}$ in (\ref{eq:sensitivity_J}) and (\ref{eq:sensitivity_R}) are still in matrix form. It can be converted into a scalar quantity using the Frobenius norm of the matrix, thereby enabling a quantitative measure of the contributions of both ${\bf{J}}\left( {\bf{\Gamma }} \right)$ and ${\bf{R}}$ to the theoretical Cramér-Rao bound ${{\bf{P}}\left( {\bf{\Gamma }} \right)}$.
\begin{equation}
{W_{\bf{J}}} = \left\| {\frac{{\partial tr\left( {{\bf{P}}\left( {\bf{\Gamma }} \right)} \right)}}{{\partial {\bf{J}}\left( {\bf{\Gamma }} \right)}}} \right\|/\left( {\left\| {\frac{{\partial tr\left( {{\bf{P}}\left( {\bf{\Gamma }} \right)} \right)}}{{\partial {\bf{J}}\left( {\bf{\Gamma }} \right)}}} \right\| + \left\| {\frac{{\partial tr\left( {{\bf{P}}\left( {\bf{\Gamma }} \right)} \right)}}{{\partial {\bf{R}}}}} \right\|} \right)
\label{eq:weight_J}
\end{equation}
\begin{equation}
{W_{\bf{R}}} = \left\| {\frac{{\partial tr\left( {{\bf{P}}\left( {\bf{\Gamma }} \right)} \right)}}{{\partial {\bf{R}}}}} \right\|/\left( {\left\| {\frac{{\partial tr\left( {{\bf{P}}\left( {\bf{\Gamma }} \right)} \right)}}{{\partial {\bf{J}}\left( {\bf{\Gamma }} \right)}}} \right\| + \left\| {\frac{{\partial tr\left( {{\bf{P}}\left( {\bf{\Gamma }} \right)} \right)}}{{\partial {\bf{R}}}}} \right\|} \right)
\label{eq:weight_R}
\end{equation}
where $\left\|  \cdot  \right\|$ represents the Frobenius norm of the matrix. The weight parameters $W_{\bf{J}}$ and $W_{\bf{R}}$ describe the proportional contributions of the current ${\bf{J}}\left( {\bf{\Gamma }} \right)$ and ${\bf{R}}$ to the Cramér-Rao bound ${{\bf{P}}\left( {\bf{\Gamma }} \right)}$ respectively.

\section{The iterative optimization framework for SS-DPE receivers}

Based on the established correlation measurement model incorporating spatial coherence and gradient, along with the Cramér-Rao Bound analysis, the SS-DPE receiver's PVT estimation can be optimized using spatially sampled correlation measurements. This section details the design, covering the optimization criterion, iteration path determination, covariance inversion, and spatial sampling strategy to guide practical implementation. 

\subsection{The optimization criterion and iteration path determination}

In contrast to the CAF maximization criterion used in GS-DPE, this paper formulates an optimization criterion based on the correlation measurement model and its spatial coherence to minimize weighted squared residuals. This allows full exploitation of the PVT information embedded in all correlation values.

Based on Bayesian estimation theory \cite{MLE}, the correlation measurement equation can be written in the form of ${\bf{z}} = {\bf{h}}\left( {\bf{\Gamma }} \right) + {\bm{\nu }}$,  and ${\bm{\nu }}$ is Gaussian noise with covariance of $\bf R$. Given these conditions, the optimization criterion to minimize the weighted squared measurement residuals are expressed as
\begin{equation}
{\bf{\hat \Gamma }} = \mathop {\arg \min }\limits_{\bf{\Gamma }} {\left( {{\bf{z}} - {\bf{h}}\left( {\bf{\Gamma }} \right)} \right)^T}{{\bf{R}}^{ - 1}}\left( {{\bf{z}} - {\bf{h}}\left( {\bf{\Gamma }} \right)} \right)
\label{eq:MLE_standard_1}
\end{equation}
where the superscript ‘$\hat{\phantom{A}}$’ represents the estimation results provided by the estimator. Building upon the aforementioned theoretical foundation, we can design an iterative estimation optimization framework that adopts the spatial coherence gradient characteristics from correlation domain perspective.

Assuming the current PVT estimation for iteration is ${{\bf{\hat \Gamma }}}_0$. It is necessary to compute the iteration direction vector ${{\bf{d}}_{\bf{\Gamma }}}$ to determine the next PVT estimation point. Ultimately, after multiple iterations, the iterative path of the PVT parameters during the optimization process will be determined until the converge criterion is met.

Specifically, the approximate Hessian matrix ${\bf{H}}$ can be derived using current Jacobian matrix ${\bf{J}}\left( {{{{\bf{\hat \Gamma }}}_0}} \right)$ and the pseudo-inverse ${{\bf{\widehat R}}^{ - 1}}$ of noise covariance matrix.
\begin{equation}
{\bf{H}} = {{\bf{J}}^T}\left( {{{{\bf{\hat \Gamma }}}_0}} \right){{\bf{\widehat R}}^{ - 1}}{\bf{J}}\left( {{{{\bf{\hat \Gamma }}}_0}} \right)
\label{eq:Hessian}
\end{equation}
where the superscript ‘$\widehat{\phantom{A}}$’ denotes the modified noise matrix. The modification is applied to avoid singularity during matrix inversion and will be discussed later.

The optimization gradient ${\bf{g}}$ of the cost function (\ref{eq:MLE_standard_1}) can be expressed using weighted measurement residuals.
\begin{equation}
{\bf{g}} = {{\bf{J}}^T}\left( {{{{\bf{\hat \Gamma }}}_0}} \right){{\bf{\widehat R}}^{ - 1}}\left( {{\bf{z}} - {\bf{h}}\left( {{{{\bf{\hat \Gamma }}}_0}} \right)} \right)
\label{eq:Gradient}
\end{equation}
where  ${{\bf{z}} - {\bf{h}}\left( {\bf{\Gamma }} \right)}$ represents the measurement residuals. Finally, the iterative direction vector ${{\bf{d}}_{\bf{\Gamma }}}$ can be obtained, thereby determining the iteration path of the PVT parameters ${\bf{\Gamma }}$.
\begin{equation}
{{\bf{d}}_{\bf{\Gamma }}} =  - {\left( {{\bf{H}} + \mu {\bf{I}}} \right)^{ - 1}}{\bf{g}}
\label{eq:Step_size}
\end{equation}
Here, $\mu$ is a damping factor that balances optimization efficiency and stability \cite{L-M optimization}. A large $\mu$ makes the optimizer approximate the Gauss-Newton method, while a small $\mu$ shifts it toward the steepest descent approach. The sequence of direction vectors ${\mathbf{d}_{\boldsymbol{\Gamma}}}$ determines the convergence path of the PVT estimation.

\subsection{Pseudo-Inverse solution for noise covariance matrix}

The implementation process in (\ref{eq:Hessian}) and (\ref{eq:Gradient}) requires inverting the noise covariance matrix $\mathbf{R}$. However, the spatial coherence among sampling points typically renders $\mathbf{R}$ singular, making direct inversion unreliable. To resolve this, an efficient and stable pseudo-inverse algorithm for $\mathbf{R}$ must be designed.

Considering the block-diagonal structure of the noise covariance matrix ${\bf{R}}$, we can first compute the pseudo-inverse of each individual block ${{\bf{R}}^m}$. To ensure the positive definiteness of matrix ${{\bf{R}}^m}$, a small quantity $\varepsilon $ is added to its diagonal elements for regularization.
\begin{equation}
{{\bf{\widehat R}}^m} = {{\bf{R}}^m} + \varepsilon {\bf{I}}
\label{eq:R_revise_single}
\end{equation}
For the positive definite matrix ${{\bf{\widehat R}}^m}$, its inverse matrix can be efficiently computed using Cholesky decomposition.
\begin{equation}
{{\bf{\widehat R}}^m} = {{\bf{W}}^m}{\left( {{{\bf{W}}^m}} \right)^T}
\label{eq:R_cholesky}
\end{equation}
\begin{equation}
{\left( {{{{\bf{\widehat R}}}^m}} \right)^{ - 1}} = {\left[ {{{\left( {{{\bf{W}}^m}} \right)}^{ - 1}}} \right]^T}{\left( {{{\bf{W}}^m}} \right)^{ - 1}}
\label{eq:R_pseudo_inverse_single}
\end{equation}
where the matrix ${{\bf{W}}^m}$ is a lower triangular matrix, allowing for accurate inversion. The final pseudo-inverse ${{\bf{\widehat R}}^{ - 1}}$ for noise covariance matrix can be expressed as a combination of the pseudo-inverses of its block-diagonal components ${\left( {{{{\bf{\widehat R}}}^m}} \right)^{ - 1}}$.
\begin{equation}
{{\bf{\widehat R}}^{ - 1}} = blkdiag\left( {\left[ {\begin{array}{*{20}{c}}
{{{\left( {{{{\bf{\widehat R}}}^1}} \right)}^{ - 1}}}&{{{\left( {{{{\bf{\widehat R}}}^2}} \right)}^{ - 1}}}& \cdots &{{{\left( {{{{\bf{\widehat R}}}^M}} \right)}^{ - 1}}}
\end{array}} \right]} \right)
\label{eq:R_pseudo_inverse_all}
\end{equation}
The pseudo-inverse ${{\bf{\widehat R}}^{ - 1}}$ of (\ref{eq:R_pseudo_inverse_all}) is subsequently used to calculate Hessian matrix ${\bf{H}}$ and optimization gradient ${\bf{g}}$ as shown in (\ref{eq:Hessian}) and (\ref{eq:Gradient}).

\subsection{The selection for spatial sparse sampling points}

The above analysis of correlation measurements, spatial coherence, spatial gradient, and optimization criterion all relies on a set of spatial sampling points ${{\bf{\tilde \Gamma }}_1},{{\bf{\tilde \Gamma }}_2}, \cdots ,{{\bf{\tilde \Gamma }}_K}$. The selection of these spatial sampling points directly determines the spatial coherence and gradients, and in turn, impacts how PVT estimation is performed. Therefore, the selection of spatial sampling points constitutes a critical component in achieving PVT parameters estimation.

Specifically, the sampling can be collected along each PVT parameters linearly, then combine all sampled dimensions to ultimately generate the full set of spatial sampling points. Considering the decoupling characteristics of position-clock bias $\left[ {{\bf{r}},\delta t} \right]$ and velocity-clock drift $\left[ {{\bf{v}},\delta \dot t} \right]$, the total spatial sampling points for $\left[ {{\bf{r}},\delta t} \right]$ dimension can be expressed as
\begin{equation}
\begin{aligned}
&\left\{ \tilde{\boldsymbol{\Gamma}}_1, \ldots, \tilde{\boldsymbol{\Gamma}}_K \right\} \\= 
&\left\{ \left( \tilde{x}, \tilde{y}, \tilde{z}, \delta\tilde{t}, \bar{\mathbf{v}}, \delta\bar{\dot{t}} \right) 
\ \middle| \ 
\tilde{x} \in S_1,\ \tilde{y} \in S_2,\ \tilde{z} \in S_3,\ \delta\tilde{t} \in S_4 \right\}
\end{aligned}
\label{eq:grid_selection_1}
\end{equation}
\begin{equation}
{S_u} = \left\{ {{\Gamma _u}\left| {{\Gamma _u} \in \left\{ {{{\bar \Gamma }_u} - \frac{{{\alpha _u}}}{2}{\beta _u}, \cdots ,{{\bar \Gamma }_u}, \cdots ,{{\bar \Gamma }_u} + \frac{{{\alpha _u}}}{2}{\beta _u}} \right\}} \right.} \right\}
\label{eq:grid_selection_2}
\end{equation}
where ${{\Gamma _u}}$ is the $u$-th element for PVT parameters ${\bf{\Gamma }}$, the overline superscript "$-$" denotes the search center, typically selected as the estimation result from the previous epoch. The search parameter $\alpha$ is a positive even integer representing the sampling intervals number, $\beta$ denotes the sampling step size, and thus $\left[ { - \alpha \beta /2,\alpha \beta /2} \right]$ represents the sampling range for this dimension. These sampling parameters collectively determine key characteristics of the SS-DPE receiver, thereby providing a reference for selecting appropriate parameters based on desired system characteristics. For clarity, we adopt the subscript “SS” for sampling parameters of the SS-DPE receiver and “GS” for search parameters of the GS-DPE receiver.

{\it Information Diversity}: The diversity of PVT information contained in the spatial sampling points is predominantly governed by the sampling step size $\beta_{SS}$. A larger sampling step size $\beta_{SS}$ reduces the singularity of the spatial coherence matrix $\mathbf{R}$, thereby promoting greater diversity in the PVT information captured by spatial sampling points. In SS-DPE receivers, the typical sampling step size is $\beta_{SS} = 150m$, whereas GS-DPE receivers employ a search step size of $\beta_{GS} = 2m$, as shown in Table. \ref{tab:Efficiency comparison}. This two-order-of-magnitude larger step size $\beta_{SS}$ reflects the essence of sparse sampling in SS-DPE framework.

\begin{table} [ht]
    \centering
         \caption{The sampling/search parameters and computation load comparison between SS-DPE and GS-DPE}
\begin{tabular}{|c|c|}
        \hline
        SS-DPE & GS-DPE \\
        \hline
        Sampling intervals number & Search intervals number \\
        $\alpha_{SS} = 2$ & $\alpha_{GS} = 150$ \\
        \hline
        Sampling step size & Search step size \\
        $\beta_{SS} = 150\text{m}$ & $\beta_{GS} = 2\text{m}$ \\
        \hline
        Number of sampling points  in x-y& Number of search points in x-y\\
        $K_{SS}=(\alpha_{SS} + 1)^2=9$ & $K_{GS}=(\alpha_{GS} + 1)^2=22801$ \\
        \hline
        Total number of sampling points & Total number of search points \\
        $K_{SS}=2(\alpha_{SS} + 1)^4=162$ & $K_{GS}=2(\alpha_{GS} + 1)^4=1.04 \times 10^9$ \\
        \hline
         Total number of correlation values&Total number of correlation values\\
         $K_{SS}M=648$ ($M=4$)&$K_{GS}M=4.16 \times 10^9$ ($M=4$)\\
         \hline
    \end{tabular}
    \label{tab:Efficiency comparison}
\end{table}

{\it Computation load}: The computation load is primarily determined by the number of correlation values to be computed, which in turn is directly dictated by sampling intervals number $\alpha_{SS}$. Given the two 4-dimensional spaces—position-clock bias $[\mathbf{r}, \delta t]$ and velocity-clock drift $[\mathbf{v}, \delta \dot{t}]$—the total number of spatial sampling points is $K_{SS} = 2(\alpha_{SS} + 1)^4$. Since all $M$ satellite channels share the same set of points, the total number of required correlation values is $K_{SS} M$. Finally, the computation load of the SS-DPE receiver and its comparison with the GS-DPE receiver are presented in Table. \ref{tab:Efficiency comparison}. 
The sparse sampling in SS-DPE enables a significant reduction in sampling intervals number $\alpha_{SS}$ compared to GS-DPE. This translates directly into a much lower number of required correlation values and computational load, while significantly boosting PVT information utilization efficiency.

{\it Convergency}: The estimation convergency of SS-DPE receivers is linked to the Cramér-Rao Bound. Furthermore, the CRB expression given in (\ref{eq:CRB_bound}) is governed by spatial coherence matrix $\mathbf{R}$ and spatial gradient matrix $\mathbf{J}(\boldsymbol{\Gamma})$, both of which are modulated by the sampling parameters $\alpha_{SS}$ and $\beta_{SS}$. Specifically, increasing sampling intervals number $\alpha_{SS}$ enlarges the sampling set, thereby raising the dimension of  spatial gradient $\mathbf{J}(\boldsymbol{\Gamma})$ and elevating the CRB. Conversely, a larger sampling step size $\beta_{SS}$ can cause spatial gradient $\mathbf{J}(\boldsymbol{\Gamma})$ at some points to diminish, which would degrade the CRB and potentially trigger divergence. Additionally, the sampling step size influences the singularity of the spatial coherence matrix $\mathbf{R}$, thereby affecting both the CRB performance and the system’s convergency behavior.

\begin{table}[ht]
    \centering
             \caption{Correspondence between system characteristics, sampling parameters, and spatial properties}
    \begin{tabular}{|c|c|c|}\hline
         System Characteristics&  Sampling Parameters& Spatial Properties\\\hline
         Information Diversity&  $\beta_{SS}$&  $\mathbf{R}$\\\hline
         Computation load&  $2(\alpha_{SS} + 1)^4$& $\dim \left( {{\bf{J}}\left( {\bf{\Gamma }} \right)} \right)$\\\hline
         Convergency&  $\alpha_{SS}$, $\beta_{SS}$& ${\bf{J}}\left( {\bf{\Gamma }} \right)$, $\mathbf{R}$\\ \hline
    \end{tabular}
    \label{tab:System characteristics}
\end{table}

In summary, the above system characteristics are governed by the spatial coherence $\mathbf{R}$ and spatial gradient ${\bf{J}}\left( {\bf{\Gamma }} \right)$, which are, in turn, fundamentally determined by the underlying spatial sampling parameters $\alpha_{SS}$ and $\beta_{SS}$, as summarized in Table. \ref{tab:System characteristics}. It needs to be noted that the similar spatial sampling strategy and system characteristics analysis is also applicable for velocity-clock drift $\left[ {{\bf{v}},\delta \dot t} \right]$, which will not be reiterated.

\subsection{Implementation process for iterative optimization}

\begin{algorithm}
    \caption{Iterative optimization steps for SS-DPE receivers}
    \renewcommand{\algorithmicrequire}{\textbf{Input:}}
    \renewcommand{\algorithmicensure}{\textbf{Output:}}
    \begin{algorithmic}[1]
        \REQUIRE IF signal points $s\left[ n \right]$, PVT parameters ${\bf{\Gamma }}^m$ for satellites, spatial sampling parameters $\alpha_{SS}$ and $\beta_{SS}$, initial PVT estimation parameters ${\bf{\hat \Gamma }}_0$, a series of factors for the optimizer (a. pseudo-inverse revision quantity $\varepsilon $; b. initial damping factor $\mu$; c. max iteration times $iter_{max}$; d. iterative step size threshold $tol$.)
        \ENSURE PVT estimation results ${\bf{\hat \Gamma }}$
        \STATE Determine the spatial sampling points $\left\{ {{\bf{\tilde \Gamma }}_1^m, \cdots ,{\bf{\tilde \Gamma }}_K^m} \right\}$ using ${\bf{\hat \Gamma }}_0$ and (\ref{eq:grid_selection_1}) - (\ref{eq:grid_selection_2})
        \STATE Construct local signal $\tilde s_I^m\left[ {n,{\bf{\tilde \Gamma }}} \right]$ using (\ref{eq:Local_I_define})
        \STATE Calculate correlation measurements $\Re _I^m\left( {{\bf{\tilde \Gamma }}} \right)$ using (\ref{eq:Correlation_I}) 
        \STATE Calculate equivalent code phase $\tilde \tau ^m$ and Doppler $\tilde f_d^m$ using (\ref{eq:Tao}) - (\ref{eq:Doppler}) for all spatial sampling points $\left\{ {{\bf{\tilde \Gamma }}_1^m, \cdots ,{\bf{\tilde \Gamma }}_K^m} \right\}$
        \STATE Obtain code phase distance  $\nabla \tau _{i,j}^m$ and Doppler distance $\nabla f_{di,dj}^m$ between ${{\bf{\tilde \Gamma }}_i}$ and ${{\bf{\tilde \Gamma }}_j}$
        \STATE Calculate noise covariance matrix ${\bf{R}}$ based on spatial coherence using (\ref{eq:R_single}) - (\ref{eq:R_all})
        \STATE Obtain pseudo-inverse ${{\bf{\widehat R}}^{ - 1}}$ using (\ref{eq:R_revise_single}) - (\ref{eq:R_pseudo_inverse_all})
        \WHILE{$iter < iter_{max}$ \AND $\left\| {{{\bf{d}}_{\bf{\Gamma }}}} \right\| > tol$} 
            \STATE Calculate Jacobian matrix ${\bf{J}}$ based on current  PVT parameters ${\bf{\hat \Gamma }}$ and spatial sampling points $\left\{ {{\bf{\tilde \Gamma }}_1^m, \cdots ,{\bf{\tilde \Gamma }}_K^m} \right\}$ using (\ref{eq:Jacobian_main}) - (\ref{eq:Jacobian_all})
            \STATE Obtain iterative direction vector ${{\bf{d}}_{\bf{\Gamma }}}$ using (\ref{eq:Hessian}) - (\ref{eq:Step_size})
            \STATE Update current PVT parameters ${{\bf{\hat \Gamma }}_{new}} = {\bf{\hat \Gamma }} + {{\bf{d}}_{\bf{\Gamma }}}$
            \STATE Calculate the optimization criterion function \\$\gamma ({{\bf{\hat \Gamma }}_{new}}) = {\left( {{\bf{z}} - {\bf{h}}\left( {{{{\bf{\hat \Gamma }}}_{new}}} \right)} \right)^T}{{\bf{\widehat R}}^{ - 1}}\left( {{\bf{z}} - {\bf{h}}\left( {{{{\bf{\hat \Gamma }}}_{new}}} \right)} \right)$
             \IF{$\gamma ({{\bf{\hat \Gamma }}_{new}}) < \gamma ({\bf{\hat \Gamma }})$} 
                    \STATE Accept the update of PVT parameters ${\bf{\hat \Gamma }} = {{\bf{\hat \Gamma }}_{new}}$
                    \STATE Reduce the damping factor $\mu  = \mu  / 10$
                \ELSE
                    \STATE Deny the update of PVT parameters and increase the damping factor $\mu  = \mu  \times 10$
                \ENDIF
                \STATE Update the iteration times $iter = iter + 1$
        \ENDWHILE
        \RETURN PVT estimation result ${\bf{\hat \Gamma }}$
    \end{algorithmic}
\label{alg:Iterative optimization for SS-DPE receivers}
\end{algorithm}

The iterative estimation framework for SS-DPE receivers can be implemented using the aforementioned theoretical analysis and processing techniques. The algorithmic flow of the entire optimization framework is shown in Algorithm \ref{alg:Iterative optimization for SS-DPE receivers}, where the optimization method used in the iterative process is based on the Levenberg-Marquardt (L-M) unconstrained optimization estimator. Noting the decoupling characteristics for position-clock bias $\left[ {{\bf{r}},\delta t} \right]$ and velocity-clock drift $\left[ {{\bf{v}},\delta \dot t} \right]$ \cite{PV decoupling}, the optimizations for them are performed independently.

\section{Theoretical performance analysis via simulation verification}

To ensure the effectiveness of the correlation measurement model including spatial coherence and spatial gradient characteristics, the Cramér-Rao Bound analysis, and the iterative optimization framework for SS-DPE receivers proposed in this paper, the theoretical performance analysis via Monte Carlo simulation verification is conducted. 

\subsection{Correlation measurement model analysis }

First, it is necessary to validate the correlation measurement model established in (\ref{eq:I_model}). To achieve this, IF signal sampling points in (\ref{eq:IF_signal_n}) and local signals (\ref{eq:Local_I_define}) need to be simulated, with the specific parameters as follows. The position and velocity of the receiver are set as ${L}=30^\circ$, ${\lambda}=120^\circ$, $h=100m$, ${\bf{v}} = \left[ {\begin{array}{*{20}{c}}
{10}&{10}&0\end{array}} \right] m/s$. The sampling frequency and correlation times are ${f_s} = 5MHz$ and ${T_c} = 20ms$. The noise density and the carrier-to-noise ratio for the IF signal are set to $\sigma  = 1000$ and $C/N_0 = 40 \, \text{dB-Hz}$, respectively. The satellite ephemeris is obtained from the Rinex file, which provides the PVT parameters ${\bf{\Gamma }}^m$ of the satellite.

\begin{figure}[ht]
    \centering
    \includegraphics[width=0.8\linewidth]{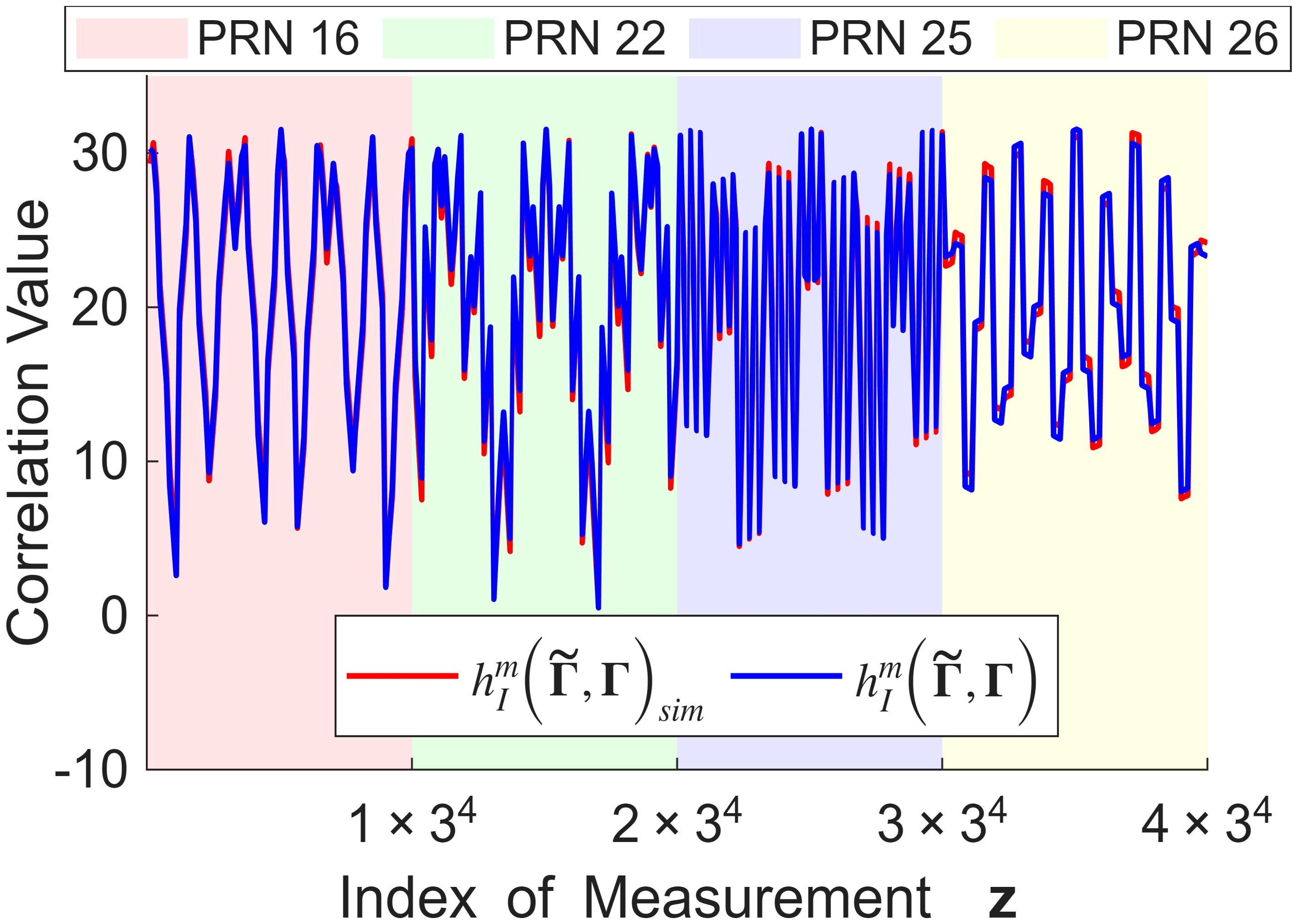}
    \caption{A comparison of the theoretical measurement model with Monte Carlo simulations demonstrates agreement across various satellites.}
    \label{fig:The measurement model}
\end{figure}

Taking the position-clock bias $\left[ {{\bf{r}},\delta t} \right]$ as an example, with the true PVT parameters $\bf{\Gamma }$ as the search center, the spatial sampling points $\bf{\tilde \Gamma }$ are determined with sampling intervals number $\alpha_{SS} = 2$ and sampling step size $\beta_{SS}=100m$. The number of satellites is $M = 4$, resulting in a total of $K_{SS} M = 4 \times 3^4$ correlation values. The simulated measurement model $h_I^m\left( {{\bf{\tilde \Gamma }},{\bf{\Gamma }}} \right)_{sim}$ is represented by correlation measurement values $\Re _I^m\left( {{\bf{\tilde \Gamma }}} \right)$ calculated using (\ref{eq:Correlation_I}), while the theoretical measurement model $h_I^m\left( {{\bf{\tilde \Gamma }},{\bf{\Gamma }}} \right)$ is calculated using (\ref{eq:Measurement_equation}). To avoid the influence of noise randomness in a single simulation, the curves in Fig. \ref{fig:The measurement model} are the averaged result of $500$ repeated Monte Carlo simulations.

It can be observed that the theoretical measurement model $h_I^m\left( {{\bf{\tilde \Gamma }},{\bf{\Gamma }}} \right)$ accurately characterizes the simulated measurement model $h_I^m\left( {{\bf{\tilde \Gamma }},{\bf{\Gamma }}} \right)_{sim}$, with nearly identical trends between the two curves. The measurement model $h_I^m\left( {{\bf{\tilde \Gamma }},{\bf{\Gamma }}} \right)$ for $\Re _I^m\left( {{\bf{\tilde \Gamma }}} \right)$ derived in (\ref{eq:I_model}) proves to be effective.

\subsection{Spatial coherence analysis}

\begin{figure}[ht]
    \centering
    \includegraphics[width=1\linewidth]{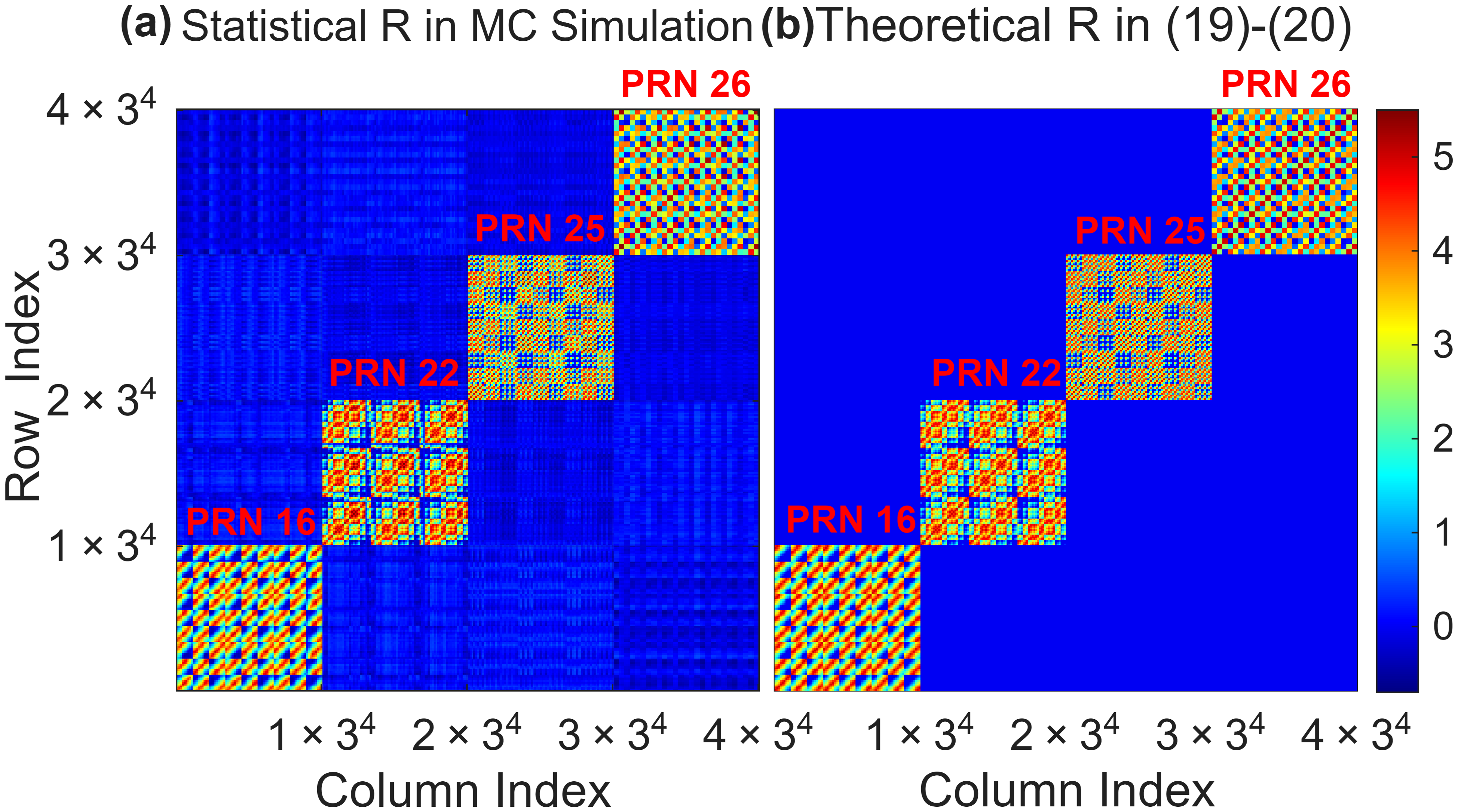}
    \caption{The noise covariance matrix $\bf R$ obtained by MC simulation and theoretical derivation. Both of them exhibit a consistent block-diagonal structure, which reflects the underlying spatial coherence.}
    \label{fig:Covariance_comparison}
\end{figure}

Still based on the above simulation parameter settings, we continue to visualize the noise covariance matrix $\bf R$ as a heatmap shown in Fig. \ref{fig:Covariance_comparison}. Fig. \ref{fig:Covariance_comparison}(a) and (b) respectively show the noise covariance matrices obtained by two methods. Specifically, Fig. \ref{fig:Covariance_comparison}(a) displays the statistical noise covariance matrix calculated from 500 repeated Monte Carlo simulations. Fig. \ref{fig:Covariance_comparison}(b) is obtained using theoretical expression of noise covariance element shown in (\ref{eq:variance_ij}) and (\ref{eq:variance_m1m2}).

The noise covariance matrix $\bf R$ in Fig. \ref{fig:Covariance_comparison}(a) exhibits a distinct block-diagonal structure, where the off-diagonal blocks represent coherence between spatial sampling points from different satellite, and their heatmap intensity is nearly negligible. In contrast, the diagonal blocks, indicating spatial coherence among different spatial sampling points from the same satellite, show significantly stronger heatmap signatures. The noise covariance matrix $\bf R$ in Fig. \ref{fig:Covariance_comparison}(b) has all its off-diagonal blocks set to zero, while the texture characteristics of its diagonal blocks are essentially consistent with those statistically obtained from Monte Carlo simulations. This agreement validates the effectiveness of the proposed theoretical calculation method for the noise covariance matrix. The distinct texture characteristics observed in the diagonal blocks of the noise covariance matrix $\bf R$ may be related to the geometric parameters of individual satellites.

\subsection{Spatial gradient analysis}

To validate the effectiveness of the derived Jacobian matrix ${\bf{J}}\left( {\bf{\Gamma }} \right)$ for the measurement function ${\bf{h}}\left( {\bf{\Gamma }} \right) $, we will compute ${\bf{J}}\left( {\bf{\Gamma }} \right)$ through both numerical method and theoretical derivation respectively, then perform comparative analysis. 

The scheme for computing the Jacobian coefficient using numerical methods is to approximate the derivatives with finite differences, and the specific approach can be written as
\begin{equation}
{{\bf{J}}^{sim}} \approx \dfrac{1}{{\delta {\bf{\Gamma }}}}\left[ {\partial h_I^m\left( {{\bf{\tilde \Gamma }},{\bf{\Gamma }} + \delta {\bf{\Gamma }}} \right) - \partial h_I^m\left( {{\bf{\tilde \Gamma }},{\bf{\Gamma }}} \right)} \right]
\label{eq:Jacobian_sim}
\end{equation}
where the selection method for the perturbation term $\delta {\bf{\Gamma }}$ is as follows: among the 8-dimensional PVT parameters $\bf{\Gamma }$, one parameter is fixed as a small perturbation, while the other 7 dimensions remain zero.
The theoretical derivation of Jacobian coefficient is calculated using (\ref{eq:Jacobian_main})-(\ref{eq:IQ_arctan2}), as well as partial derivative items in Table \ref{eq:Partial_derivative_items}.

\begin{figure}[ht]
    \centering
    \includegraphics[width=0.8\linewidth]{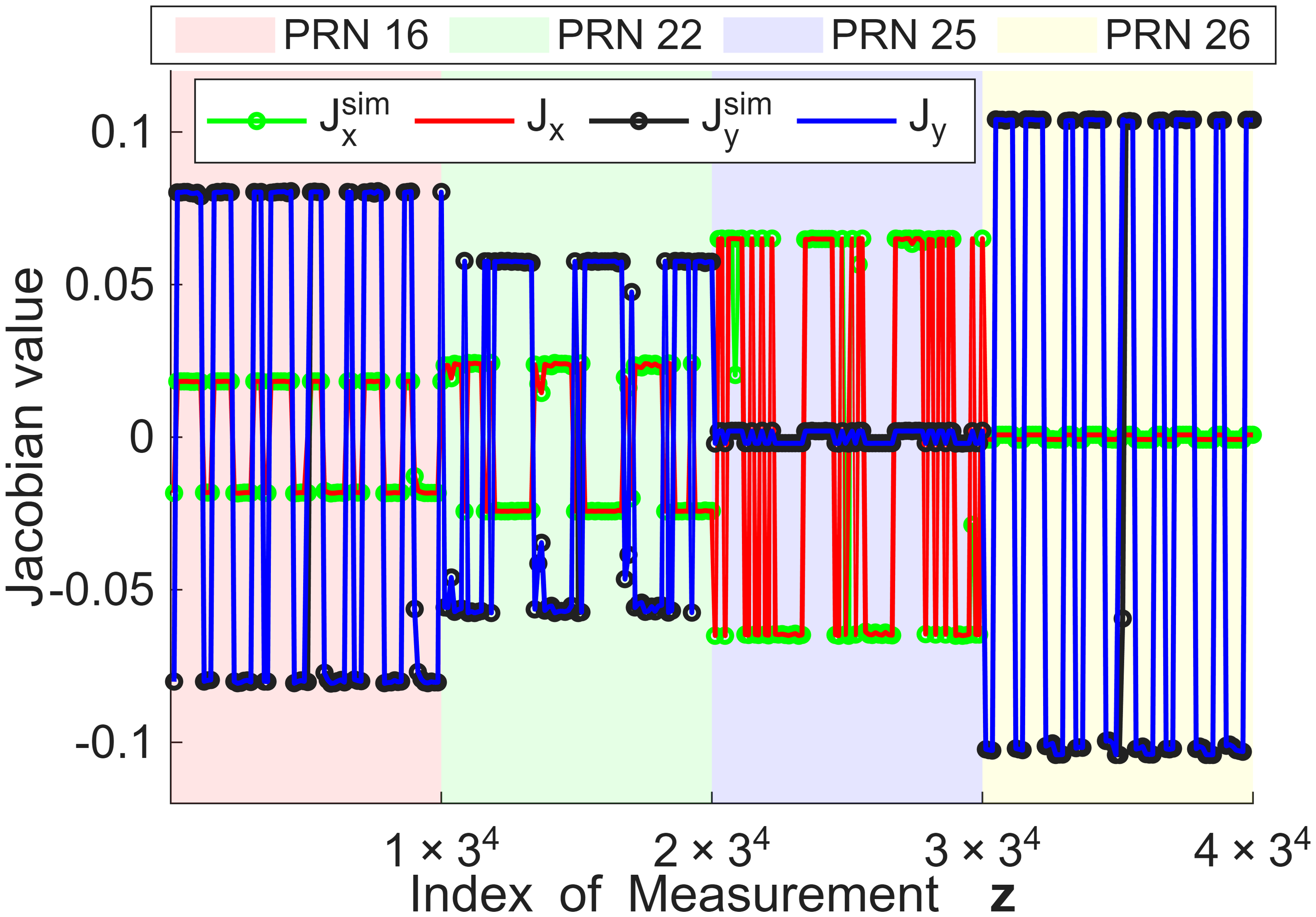}
    \caption{The comparison of Jacobian matrix ${\bf{J}}\left( {\bf{\Gamma }} \right)$ using numerical method and theoretical derivation. }
    \label{fig:Jacobian comparison}
\end{figure}

Taking the $x$ and $y$ position as an example, Fig. \ref{fig:Jacobian comparison} shows the Jacobian coefficients corresponding to all measurements $\bf z$ in Fig. \ref{fig:The measurement model}, obtained by both numerical method and theoretical derivation. In the numerical method, the perturbation terms $\delta {\bf{\Gamma }}$ are respectively set to $5m$. The Jacobian coefficients for all measurements $\mathbf{z}$, obtained through both numerical and theoretical approaches, show close agreement. This consistency validates the effectiveness of the derived coefficients. The comparison of the Jacobian coefficients for the same satellite measurement reveals that they approximately oscillate between fixed positive and negative values. This is attributed to the derivative item $\partial \Re _\tau ^m\left( {\Delta {\tau ^m}} \right)/\partial {\tau ^m}$ of (\ref{eq:Jacobian_tao}), which holds inherent oscillatory nature of the ${\rm{sig}}n\left( {\Delta {\tau ^m}} \right)$ function. Variations in the oscillatory amplitude of the Jacobian coefficients across different satellites arise mainly from differences of the geometric term $\partial \tau^m / \partial \boldsymbol{\Gamma}$ in Table \ref{eq:Partial_derivative_items}.

\subsection{The Cramér-Rao Bound analysis}

To verify the effectiveness of the Cramér–Rao bound and its sensitivity analysis, corresponding simulations are carried out. Two cases were considered: (1) the position–clock bias $\left[ {{\bf{r}},\delta t} \right]$, and (2) the velocity–clock drift  $\left[ {{\bf{v}},\delta \dot t} \right]$. Both sampling center are set at the true point, and the impact of different sampling parameters on the CRB and its sensitivity is analyzed.

\begin{figure}[ht]
    \centering
    \includegraphics[width=0.8\linewidth]{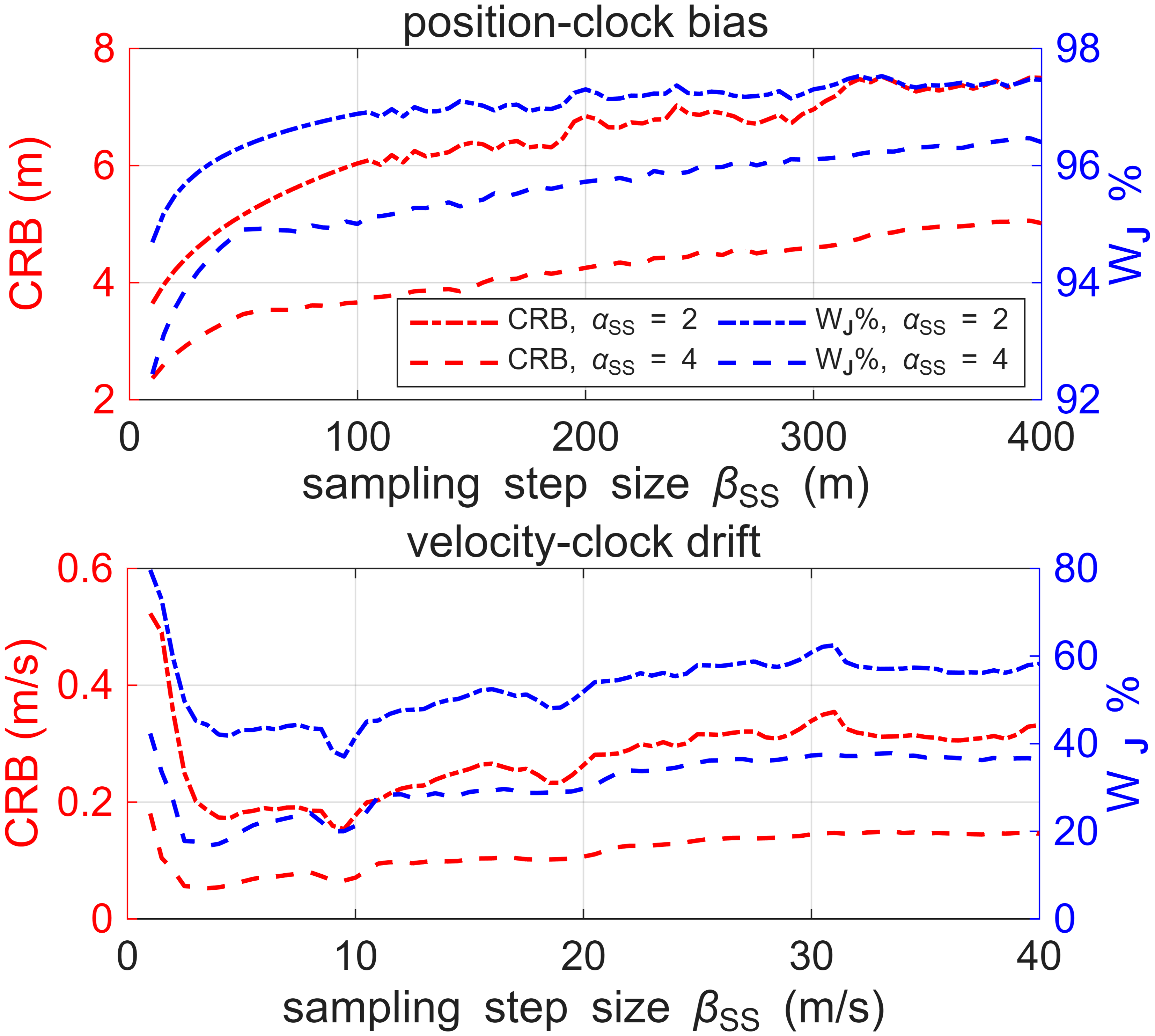}
    \caption{The CRB for PVT estimation and its sensitivity results with different sampling intervals number $\alpha_{SS}$ and sampling step size $\beta_{SS}$.}
    \label{fig:CRLB_result}
\end{figure}

As illustrated in Fig. \ref{fig:CRLB_result}, for the curve with a fixed sampling intervals number $\alpha_{SS} = 2$, the CRB exhibits a monotonically increasing trend as the sampling step size $\beta_{SS}$ grows. This occurs because sampling points farther from the true PVT point contain less PVT information, resulting the spatial gradient matrix $\mathbf{J}(\boldsymbol{\Gamma})$ with a smaller magnitude. More specifically, the CRB curve exhibits an initial smooth period followed by fluctuations. This behavior stems from the derivative item $\partial \Re _\tau ^m\left( {\Delta {\tau ^m}} \right)/\partial {\tau ^m}$ of (\ref{eq:Jacobian_tao}), where the $rect(\cdot)$ function remains constant at $1$ when the code phase error $\left| {\Delta {\tau ^m}} \right| \le 1$ and drops to $0$ when $\left| {\Delta {\tau ^m}} \right| > 1$. When sampling step size $\beta_{SS}$ is small, all spatial sampling points satisfy the condition of  $\left| {\Delta {\tau ^m}} \right| \le 1$, resulting in a smoothly increasing CRB trend. As the sampling step size $\beta_{SS}$ further increases, some spatial sampling points experience code phase errors $\left| {\Delta {\tau ^m}} \right| > 1$, causing $rect(\cdot)$ function from $1$ to $0$ and consequently fluctuations in the CRB. This phenomenon fundamentally arises from the non-uniform change rate of correlation measurements across different spatial directions.

For a fixed sampling intervals number $\alpha_{SS} = 4$, the CRB is lower than with $\alpha_{SS} = 2$. This improvement arises because, at the same step size $\beta_{SS}$, more sampling points over a wider range yield a larger set of correlation measurements, which provide richer PVT information. Furthermore, the smooth region of CRB curve for $\alpha_{SS} = 4$ is shorter, which results from the expanded sampling range causing points with $\left| {\Delta {\tau ^m}} \right| > 1$ to appear earlier. The subsequent fluctuation amplitude is also smaller, as the increased number of sampling points diminishes the impact of $rect(\cdot)$ function degeneration from $1$ to $0$ in individual points on the overall performance.

For the variation trend of CRB in the velocity–clock drift $\left[ {{\bf{v}},\delta \dot t} \right]$ as shown in Fig. \ref{fig:CRLB_result}, its most significant distinction from the position-clock bias $\left[ {{\bf{r}},\delta t} \right]$ curve lies in the following: the CRB first decreases and then increases as the sampling step size $\beta_{SS}$ grows, accompanied by local fluctuations throughout the entire process. This occurs because the derivative term $\partial \Re _f^m\left( {\Delta f_d^m} \right)/\partial f_d^m$ of (\ref{eq:Jacobian_Doppler}) initially increases and then decreases as the Doppler error ${\Delta f_d^m}$ grows inside the main peak region, while exhibiting oscillatory characteristics outside the main peak region. This behavior consequently leads to both the overall trend and local fluctuations observed in the corresponding CRB curve.

The proportion percent $W_{\bf J}\%$ of Jacobian matrix ${\bf{J}}\left( {\bf{\Gamma }} \right)$ on total CRB is also plotted in Fig. \ref{fig:CRLB_result}. The spatial gradient weight $W_{\mathbf{J}}$ exceeds $92\%$ for position-clock bias $[\mathbf{r}, \delta t]$, whereas it varies between $20\%$ - $80\%$ for velocity-clock drift $[\mathbf{v}, \delta \dot{t}]$ because the magnitude of $\partial \Re_f^m (\Delta f_d^m) / \partial f_d^m$ is inherently much smaller than $\partial \Re_\tau^m (\Delta \tau^m) / \partial \tau^m$. This insight informs the sampling strategy: the selection of spatial sampling points should be tailored according to the actual weighting ratio, emphasizing different spatial characteristics accordingly.

\begin{figure}[ht]
    \centering
    \includegraphics[width=0.8\linewidth]{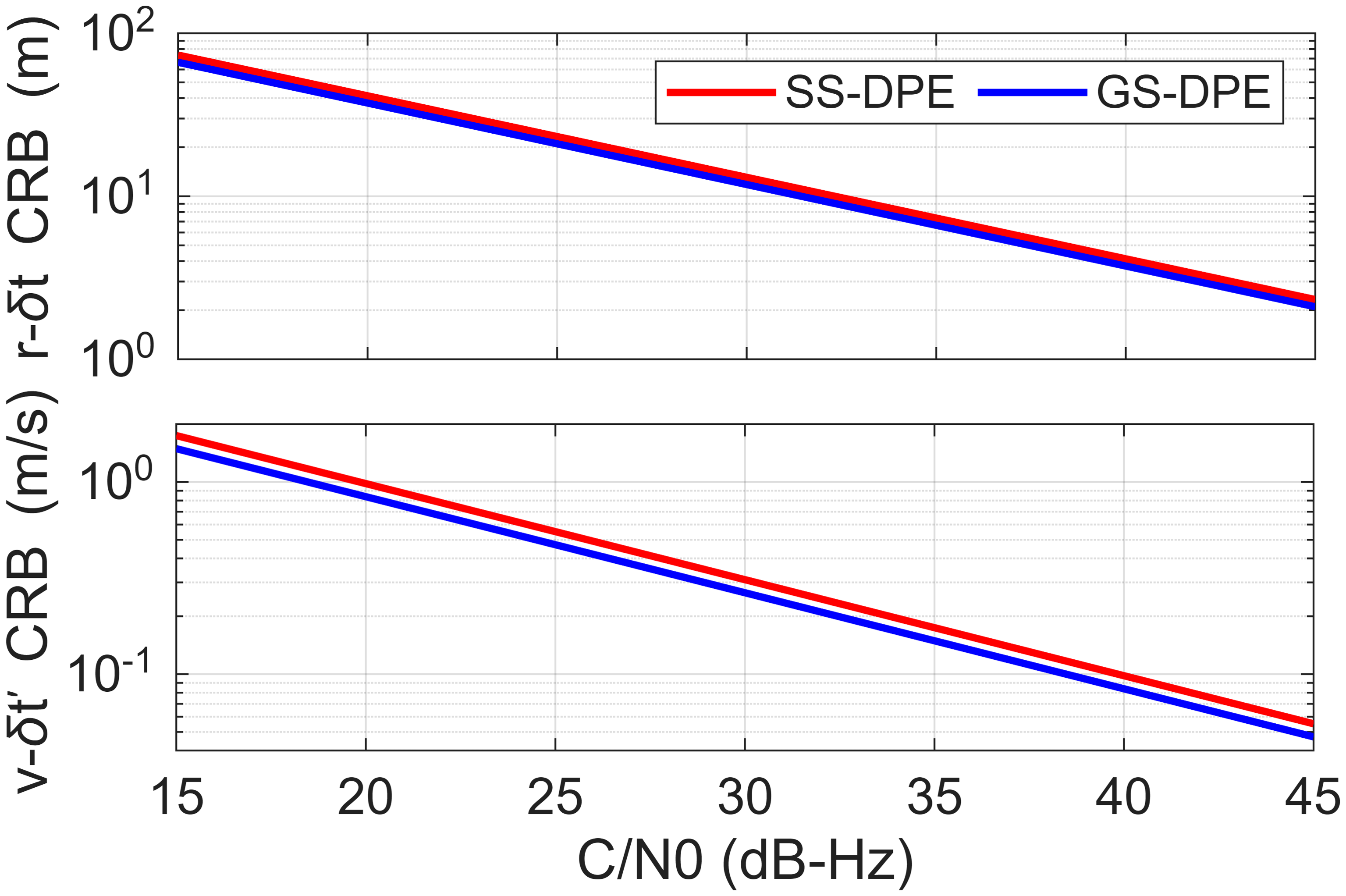}
    \caption{The CRB for PVT estimation in traditional GS-DPE and our proposed SS-DPE under different signal $C/{N_0}$. The SS-DPE achieves comparable theoretical accuracy with significantly reduced spatial sampling points.}
    \label{fig:CRLB_CN0}
\end{figure}

Using fixed spatial sampling parameters selected based on Fig. \ref{fig:CRLB_result}, the theoretical CRB for both SS-DPE and GS-DPE across varying $C/N_0$ is plotted in Fig. \ref{fig:CRLB_CN0}. The GS-DPE CRB is computed following \cite{DPE Cramer–Rao Bound}. It can be observed that the CRB curves of the two methods are very close, indicating that the theoretical performance of the SS-DPE is comparable to that of the GS-DPE. 
Under the signal $C/N_0 = 45 \, \text{dB-Hz}$, the CRB for position–clock bias $\left[ {{\bf{r}},\delta t} \right]$ of the SS-DPE receiver reaches $2.3m$, and $0.06 m/s$ for velocity–clock drift $\left[ {{\bf{v}},\delta \dot t} \right]$. The persistent, slight CRB gap between SS-DPE and GS-DPE stems from the intrinsic characteristic: as linear combinations of original IF signal points $s[n]$,  correlation values $\Re_I^m(\tilde{\boldsymbol{\Gamma}})$ implies an inherent performance limitation, bounding estimation below the level achievable by the original IF signal points $s[n]$.

\subsection{Iteration convergence and computational complexity}

\begin{figure}[ht]
    \centering
    \includegraphics[width=0.8\linewidth]{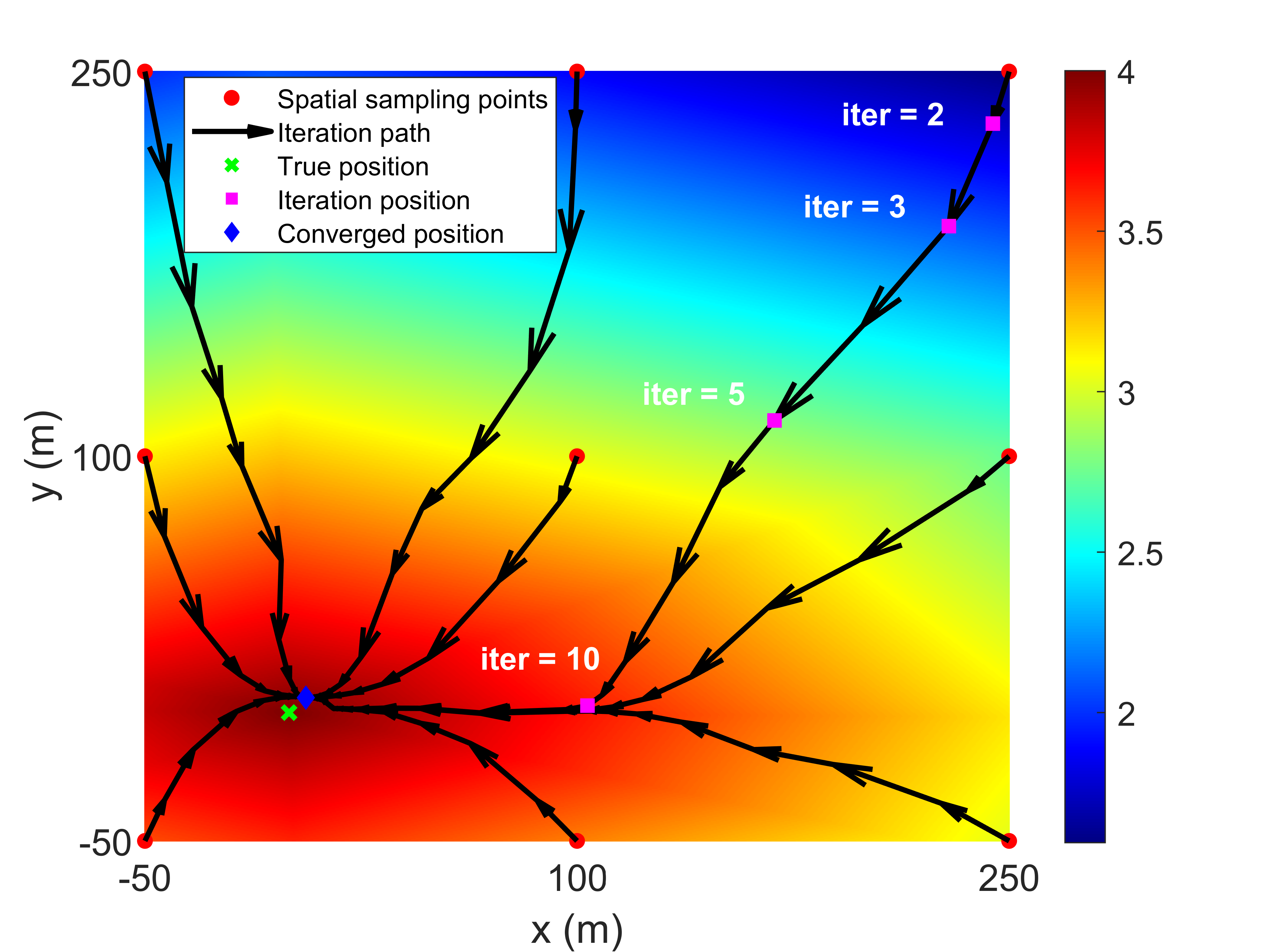}
    \caption{The convergence path and the selection of spatial sampling points for SS-DPE. Robust convergence is achieved across different initial points.} 
    \label{fig:Iteration_path}
\end{figure}

Using the process summarized in Algorithm \ref{alg:Iterative optimization for SS-DPE receivers}, the iterative optimization for SS-DPE receivers can be implemented. We chose the $x-y$ position plane for better clarity to present the details of the proposed method, as shown in Fig. \ref{fig:Iteration_path}. 

The spatial sampling points required for SS-DPE are significantly sparse. With the sampling centered at $(100m, 100m)$, sampling intervals number $\alpha_{SS} = 2$ and sampling step size $\beta_{SS}=100m$, only $K_{SS}={\left( {\alpha_{SS} + 1} \right)^2}=9$ spatial sampling points are selected across the plane, as shown in Fig. \ref{fig:Iteration_path}. These $9$ spatial points yield a total of $K_{SS}M=36$ correlation measurements (satellite channels number $M=4$). When these $9$ spatial points are individually used as initial points for SS-DPE algorithm, they all ultimately converge to the same position $(5.86m, 5.87m)$ close to true position $(0m, 0m)$. This demonstrates the feasibility, efficiency, and stability of the proposed SS-DPE algorithm.

\begin{figure}[ht]
    \centering
    \includegraphics[width=0.8\linewidth]{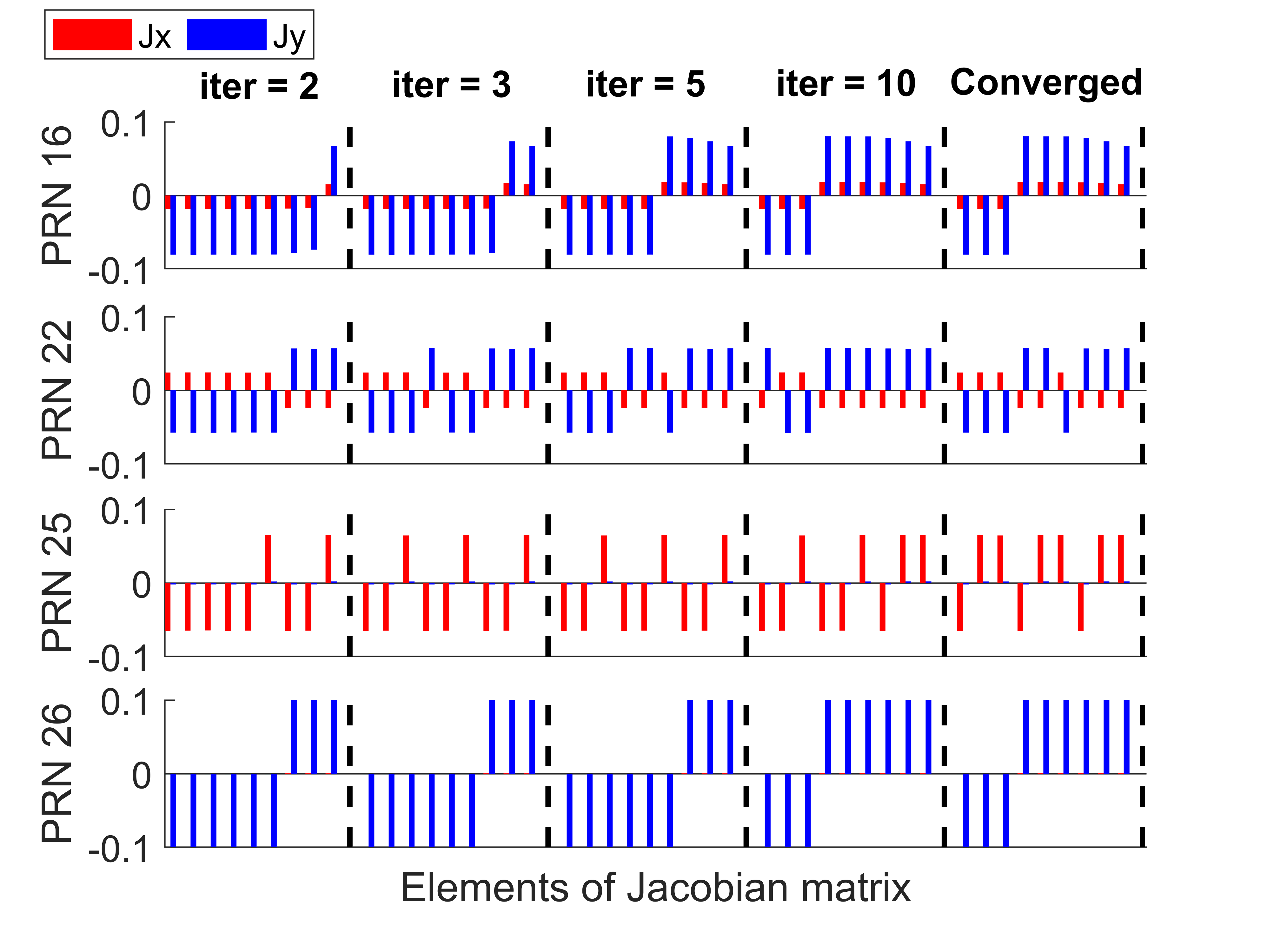}
    \caption{The convergence of the Jacobian coefficient during the iteration process. The varying Jacobian guides the direction and path of iterations.}
    \label{fig:Iteration_Jacobian}
\end{figure}

Taking the initial point at $(250m, 250m)$ for SS-DPE as an example,  Fig. \ref{fig:Iteration_Jacobian} illustrates the changes of Jacobian coefficients in both $x$ and $y$ directions during the iterative process. These intermediate points are also marked along the iteration path shown in Fig. \ref{fig:Iteration_path}. 

It can be observed that the magnitude of Jacobian coefficients for each satellite remains essentially constant throughout the iteration process, with only the sign alternating between positive and negative. This behavior aligns  with the oscillation characteristics induced by ${\rm{sig}}n\left( {\Delta {\tau ^m}} \right)$ function of the derivative item $\partial \Re _\tau ^m\left( {\Delta {\tau ^m}} \right)/\partial {\tau ^m}$ analyzed in Fig. \ref{fig:Jacobian comparison}. The variation in satellite geometry induces differences in $\partial \tau^m / \partial \boldsymbol{\Gamma}$ items, thereby introducing different Jacobian coefficients along the $x$ and $y$ directions. The Jacobian matrix, collectively constructed from all satellites and all spatial sampling points, determines both the forward direction and forward step for the current iteration point, ultimately driving the system to convergence.

\section{Experiment Verification}

To further investigate the performance of the SS-DPE algorithm proposed in this paper, this section will present the results of experiment verification in real-world scenarios.

\subsection{Experimental setup}

\begin{figure}[ht]
    \centering
    \includegraphics[width=0.7\linewidth]{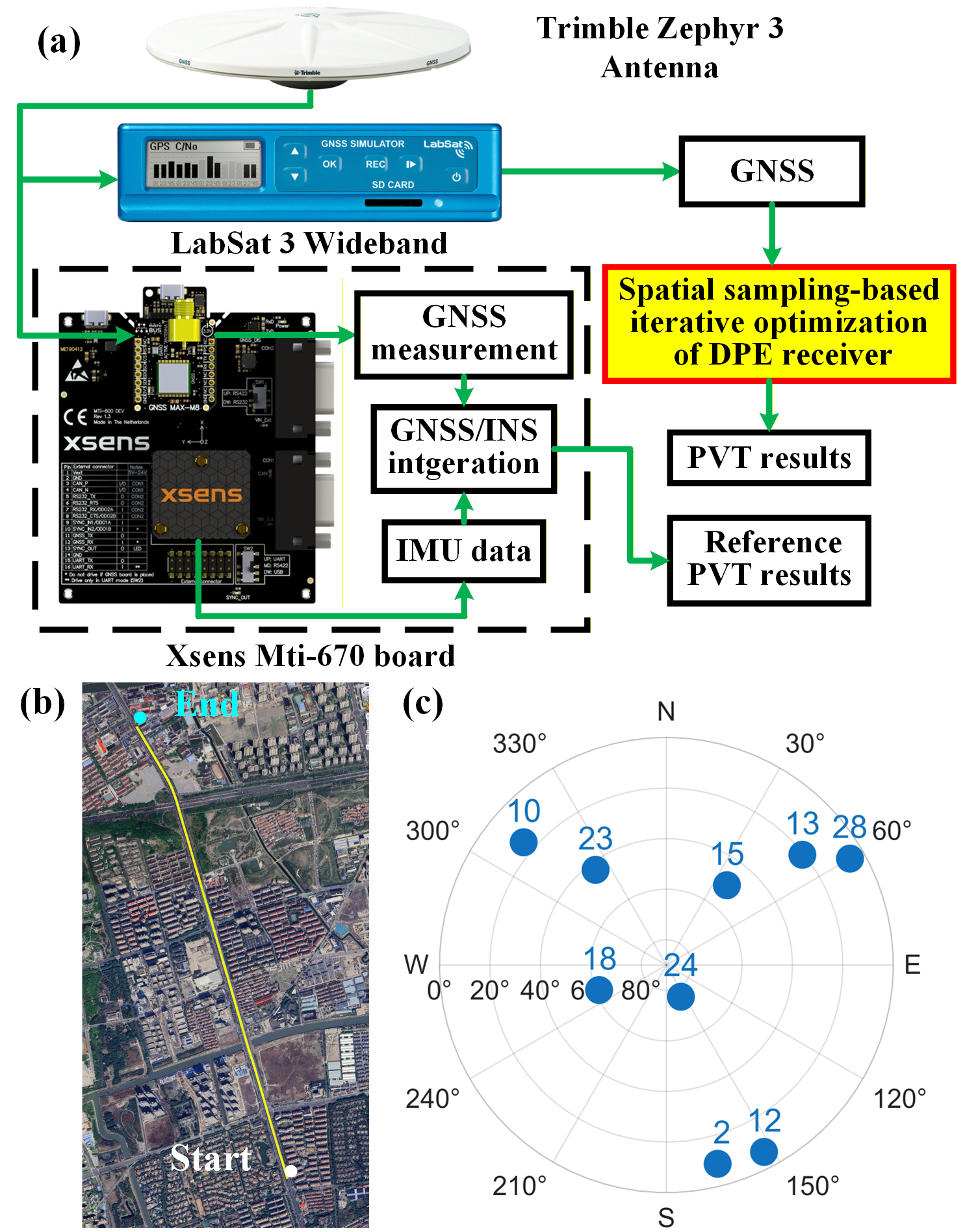}
    \caption{The hardware platform, test route, and sky plot for the vehicle test. The IF data is used to validate the algorithm, while the GNSS/INS solution serves as the reference.}
    \label{fig:equipment_route}
\end{figure}

The hardware platform used for experimental validation is shown in Fig. \ref{fig:equipment_route}(a), where a Trimble Zephyr 3 antenna is employed to capture GNSS signals, and a Labsat 3 wideband device stores IF signal data of GPS L1 C/A \cite{Labsat}. The IF signal holds a sampling frequency of $f_s=30.69MHz$ with a 2-bit quantization. In order to conduct performance validation of the proposed algorithm, the Xsens Mti-670 board is equipped to provided the reference PVT results. The Xsens board is a commercial GNSS/INS integration system, equipped with an IMU of $400Hz$ data rate and a GNSS module of multifrequency/multiconstellation u-blox Max M8Q \cite{Xsens}. It achieves a standard positioning accuracy of $1m$ and a velocity accuracy of $0.05m/s$, making it suitable as a ground reference.

The vehicle test route located at the Hongmei Elevated Road in Minhang District, Shanghai. The test was conducted on September 3, 2022, with a total valid duration of $150s$. The route trajectory is marked on Google Map, as shown in Fig. \ref{fig:equipment_route}(b). The test route is well-suited for preliminarily validating the feasibility of the SS-DPE algorithm, as it consists primarily of viaducts. The sky plot of GPS satellites in the vehicle test is shown in  Fig. \ref{fig:equipment_route}(c). We selected the same subset of satellites with good signal qualities for algorithm validation to ensure the fairness of the comparative experiments.

\subsection{PVT results analysis}

\begin{figure}[ht]
    \centering
    \includegraphics[width=0.8\linewidth]{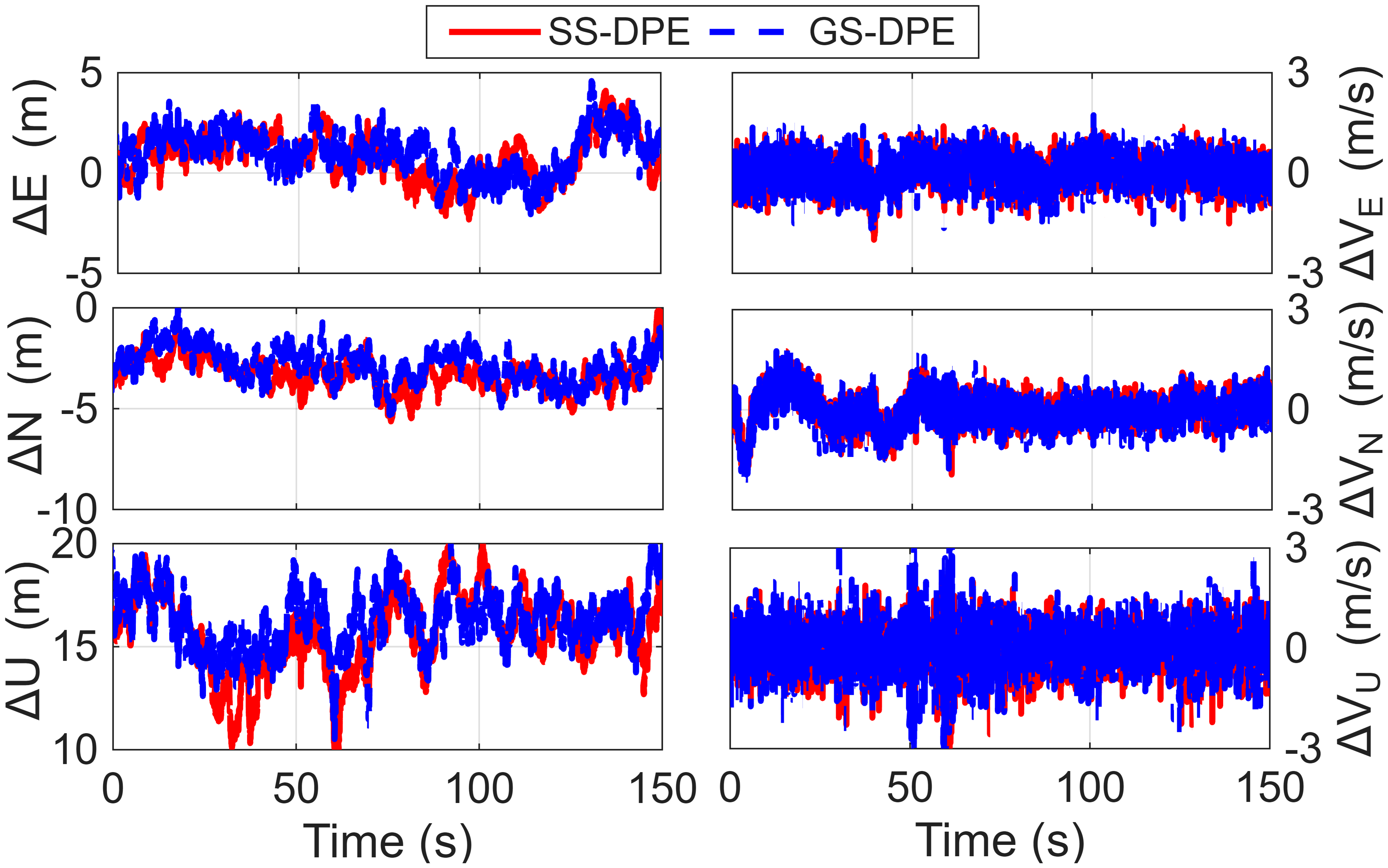}
    \caption{The position and velocity error for SS-DPE and GS-DPE during the vehicle test. The two methods deliver comparable positioning performance.}
    \label{fig:PVT_error}
\end{figure}

Fig. \ref{fig:PVT_error} presents the position and velocity error of the SS-DPE algorithm throughout the entire vehicle test, respectively. The selection of spatial sampling points in the vehicle test is optimized by incorporating both spatial coherence and spatial gradient. The results of GS-DPE algorithm are also included.

As shown in Fig. \ref{fig:PVT_error}, both the SS-DPE and GS-DPE methods exhibited position errors within $ \pm 5m$ in the horizontal directions. However, the vertical direction error showed a bias of approximately $15 m$. As for velocity, the errors are within $\pm 2 m/s$ in the horizontal directions and $\pm 3 m/s$ vertically. The observed bias and error fluctuations in the vertical direction are primarily due to poor satellite geometry and atmospheric delays. In summary, the SS-DPE and GS-DPE schemes demonstrated comparable performance under the tested conditions. Notably, the SS-DPE scheme achieved this by leveraging efficient spatial sparse sampling, which offers high information utilization efficiency with substantially reduced spatial points and computational cost.

\subsection{Iterative path analysis}

\begin{figure}[ht]
    \centering
    \includegraphics[width=0.8\linewidth]{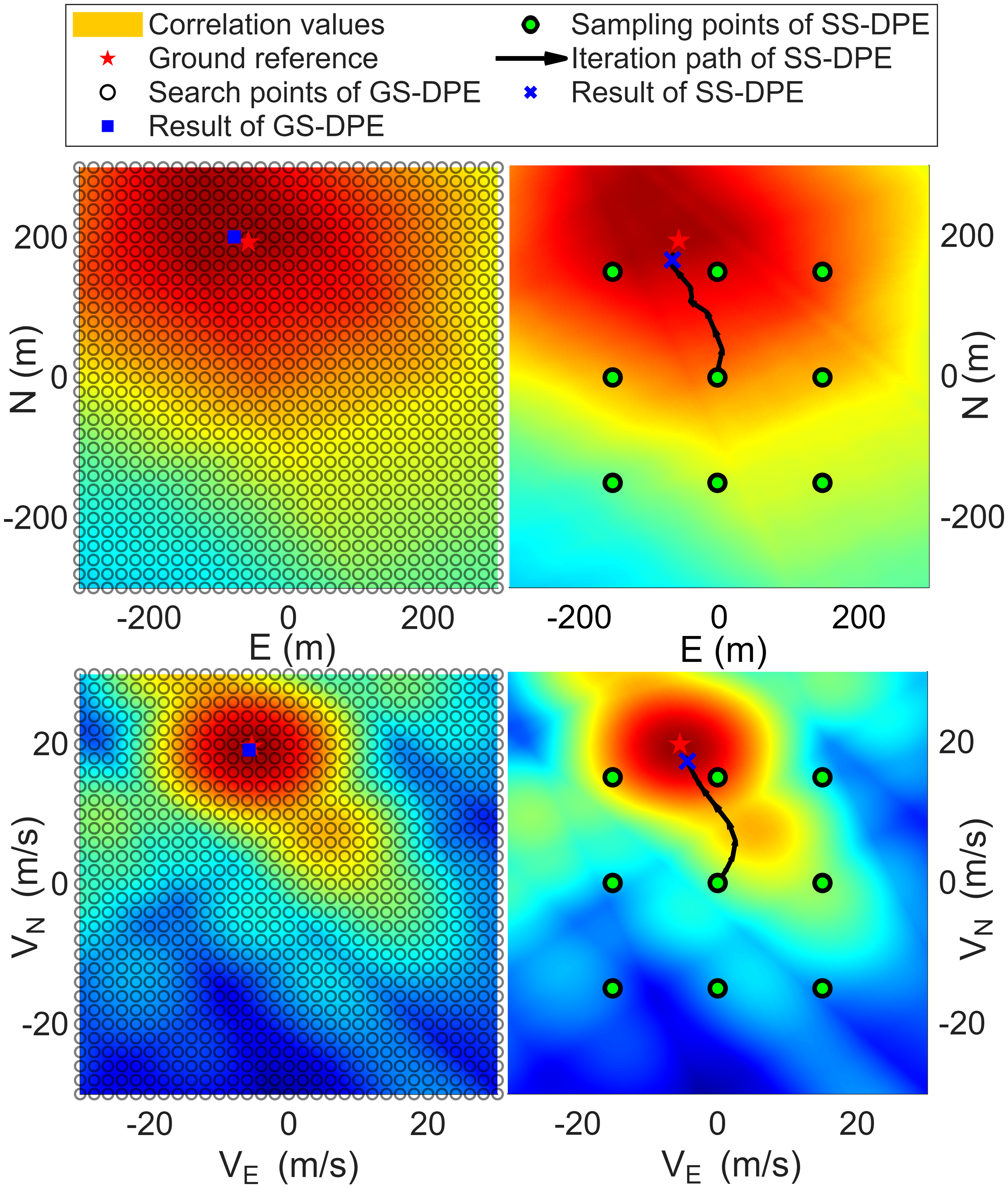}
    \caption{Taking the data at $75s$ as an example, the PVT estimation results of GS-DPE and SS-DPE are presented and compared. While GS-DPE requires covering the entire plane with search points, SS-DPE achieves similar accuracy with only $9$ sparse spatial sampling points.}
    \label{fig:grid_opt_comparison}
\end{figure}

Due to the different working principles of SS-DPE and GS-DPE, their working conditions at specific moments can be observed and compared from the perspectives of energy accumulation and iterative paths, respectively. Taking the half time point of $75s$ in the vehicle test as an example, the working conditions of the two methods can be presented in position and velocity dimensions, as shown in Fig. \ref{fig:grid_opt_comparison}.

For clarity, only the two horizontal position dimensions are shown during search or iteration, while vertical position and clock bias are held constant. The same applies to the velocity dimension. To distinguish between current position and estimated position, the current central position for search or spatial sampling is set to the reference position of time point $65s$, while the central velocity is set to $(0 m/s, 0 m/s)$. Compared to the multiple iteration paths determined by different initial iteration values in the simulation shown in Fig. \ref{fig:Iteration_path}, the experiment result of SS-DPE uses one representative path for demonstration, starting from the central sampling point.

The GS-DPE uses search intervals number of $\alpha_{GS} = 300$ and sampling step size of $\beta_{GS} = 2m$ for position, $\beta_{GS} = 0.2m/s$ for velocity (The distance between adjacent search points shown in Fig. \ref{fig:grid_opt_comparison} is $20m$ for position and $2m/s$ for velocity, thus avoiding overlap and ensuring visual clarity of the figure). Due to the dense search grid, the total number of search points in GS-DPE is $K_{GS} = (\alpha_{GS} + 1)^2 = 90601$. This high density results in these points nearly covering the entire plane shown in the left part of Fig. \ref{fig:grid_opt_comparison}. The SS-DPE method employs the sampling intervals number of $\alpha_{SS} = 2$ and sampling step size of $\beta_{SS} = 150m$ for position, $\beta_{SS} = 15m/s$ for velocity, resulting in total $K_{SS}={\left( {\alpha_{SS} + 1} \right)^2}=9$ spatial sampling points in the horizontal plane. They are presented by $9$ circles in the right part of Fig. \ref{fig:grid_opt_comparison}.

It can be seen that the energy peaks for GS-DPE align well with the ground reference, and the iteration paths for SS-DPE also converge rapidly to the ground reference. Both GS-DPE and SS-DPE can achieve the PVT estimation task with relatively high accuracy. In contrast to the GS-DPE method, whose correlation values covers the entire plane, the SS-DPE employs a sparse set of only 9 sptial sampling points. Moreover,the SS-DPE receiver achieves convergence even when the energy peak falls outside the sampling range, demonstrating the advantage of the proposed method. This stark contrast demonstrates the high information utilization efficiency and less computation load of the SS-DPE approach.
 
\section{Conclusion}

This paper establishes a spatial sparse sampling-based iterative optimization DPE framework (SS-DPE). In contrast to conventional GS-DPE, which discards the majority of correlation values, the SS-DPE framework fully utilizes the complete set of correlation values as effective measurements for PVT estimation. The spatial coherence matrix of these correlation measurements exhibits a characteristic block-diagonal structure, while the spatial gradient displays an oscillatory pattern with alternating signs. These two spatial properties collectively determine the theoretical CRB for PVT estimation. Under the signal quality of $C/N_0 = 45 \, \text{dB-Hz}$, the SS-DPE framework achieves a theoretical positioning CRB of $2.3m$ for position–clock bias and $0.06m/s$ for velocity–clock drift, matching the performance of GS-DPE under identical conditions. Moreover, in real-world vehicle test, the SS-DPE receiver, utilizing only $9$ sparse spatial sampling points, attained a position accuracy of $5m$ and velocity accuracy of $2m/s$. This performance is comparable to that of the GS-DPE receiver, while reducing the required number of correlation value computations to less than $0.05\%$ of the latter. These results conclusively demonstrate that the proposed SS-DPE framework achieves high information utilization efficiency and a drastically reduced computation load, validating its potential for efficient and robust PVT estimation.

Future research on SS-DPE should focus on providing theoretical support to further optimize spatial sampling strategies for reconstructing PVT information with as few spatial sampling points as possible. Notably, compared to the conventional CAF maximization criterion, the current optimization criterion to minimize weighted squared residuals offers greater advantages for multi-sensor fusion. The future effort can be devoted to the investigation of GNSS DPE integration with the different navigation sensors under the proposed iterative optimization framework, which should be particularly significant and promising.

\vfill

\end{document}